\documentclass[namedreferences]{solarphysics}

\usepackage[hyperref,optionalrh]{spr-sola-addons} 
\usepackage{graphicx}        
\usepackage{amssymb}        
\usepackage{color}           
\usepackage{graphicx}

\newcommand{\etal}{{\it \etal}}

\chardef\us=`\_

\begin{document}

\begin{article}
\begin{opening}

\title{CME Dynamics Using STEREO and LASCO Observations: The Relative Importance of Lorentz Forces and Solar Wind Drag}

\author[addressref={aff1},corref,email={nishtha.sachdeva@students.iiserpune.ac.in}]{\inits{Nishtha}\fnm{N.}~\lnm{Sachdeva}}
\author[addressref={aff1,aff2}]{\inits{Prasad}\fnm{P.}~\lnm{Subramanian}}
\author[addressref={aff3,aff4}]{\inits{Angelos}\fnm{A.}~\lnm{Vourlidas}}
\author[addressref={aff5}]{\inits{Volker}\fnm{V.}~\lnm{Bothmer}}

\address[id=aff1]{Indian Institute of Science Education and Research, Dr. Homi Bhabha Road, Pashan, Pune - 411008, India}
\address[id=aff2]{Centre for Excellence in Space Sciences, Kolkata India, http://www.cessi.in}
\address[id=aff3]{The Johns Hopkins University, Applied Physics Laboratory, Laurel, MD 20723 USA}
\address[id=aff4]{IAASARS, National Observatory of Athens, GR-15236, Penteli, Greece}
\address[id=aff5]{University of G\"{o}ttingen, Institute of Astrophysics, Friedrich-Hund-Platz 1, 37077 G\"{o}ttingen, Germany}
\runningauthor{Sachdeva {\it et al.}}
\runningtitle{CME dynamics}

\begin{abstract}
We seek to quantify the relative contributions of Lorentz forces and aerodynamic drag on the propagation of solar coronal mass
ejections (CMEs).
We use Graduated Cylindrical Shell (GCS) model fits to a representative set of 38 CMEs observed with the \textit{Solar and Heliospheric Observatory}
(SOHO) and \textit{Solar and TErrestrial RElations Observatory} (STEREO) spacecraft. 
We find that the Lorentz forces generally peak between 1.65 and 2.45 R$_{\odot}$ for all CMEs.
For fast CMEs, Lorentz forces become negligible in comparison to aerodynamic drag as early as 
3.5\,--\,4 R$_{\odot}$. For slow CMEs, however, they become negligible only by 12\,--\,50 R$_{\odot}$. For these slow events, our 
results suggest that some of the magnetic flux might be expended in CME expansion or heating. In other words, not all of it
contributes to the propagation.
Our results are expected to be important 
in building a physical model for understanding the Sun\,--\,Earth dynamics of CMEs.
\end{abstract}

\end{opening}

\section{Introduction}
     \label{S-Introduction} 
Coronal mass ejections (CMEs) from the Sun are generally acknowledged as the main cause of disturbances 
in the near-Earth space environment. Due to the considerable technological impacts caused by such disturbances,
it has become increasingly important to study and understand various aspects related to CME impacts on the 
Earth's magnetosphere \citep{Gos91,Bot07}. One of the most basic quantities in this regard concerns the time it takes for a CME to reach 
the Earth after it has been detected leaving the Sun using space-based coronagraphs; This quantity is called 
the Sun\,--\,Earth travel time.  Among the factors that affect the travel time are the CME size, 
mass, initial velocity and the ambient solar wind speed \citep{Bos12}. 
An accurate and reliable forecast of the Sun\,--\,Earth travel time is obviously important to a space weather 
mitigation framework, as is a good estimate of the expected speed of the CME near the Earth. These quantities are 
typically computed from a dynamical model for CME propagation, that uses near-Sun coronagraph
observations as input.

The basic outlines of such dynamical CME propagation models have been well established for a while. One dimensional (1D) models 
that incorporate Lorentz force driving, aerodynamic drag, and other effects have been in vogue since 1996 
\citep[\textit{e.g.}][]{Che96,Kum96}. So-called drag-based models (DBM), which consider 
only aerodynamic drag, have been very popular lately 
(\textit{e.g.} \citealp{Car04}; \citealp{Vrs10}; 
 \citealp{Mis13}; \citealp{Tem15}). 
More sophisticated three dimensional (3D) MHD models such as ENLIL \citep[\textit{e.g.}][]{Tak09,Lee13,Vrs14,May15}, global MHD model using the data-driven Eruptive Event 
Generator Gibson-Low (EEGGL) \citep[\textit{e.g.}][]{Jin17}, CME and shock propagation models like the Shock Time of Arrival Model (STOA), the Interplanetary 
Shock Propagation model (ISPM) and the Hakamada-Akasofu-Fry version 2 model 
\citep[\textit{e.g.}][]{Fry03,McK06}, other hybrid models \citep{Wu07} and the Space Weather 
Modeling Framework \citep[SWMF;][]{Lug07,Tot07} are also often used in modeling CME propagation.

Despite considerable progress, our ability to successfully model the Sun\,--\,Earth travel time and the near-Earth 
speed of Earth-directed CMEs is still limited \citep{Zha14}, even for relatively simple events that do not involve 
interacting CMEs \citep[\textit{e.g.}][]{Tem12}. Part of the reason for this is that the models are still largely empirical. For instance, most drag-based models use the 
dimensionless drag coefficient $C_{\rm D}$ and/or the parameter $\gamma$ (ratio of drag acceleration and square of the difference between the CME and solar wind speeds) as a fitting parameter. The physical basis of the aerodynamic 
drag experienced by CMEs is only starting to be understood \citep{Sub12,Sac15}. As far as Lorentz forces go, it is also generally thought 
that they are dominant only in the initial phases of CME propagation, when they are relatively near the Sun. However, for a given CME, its not clear
where Lorentz forces peak and when they cease to be important. Some 1D models 
(\textit{e.g.} \citealp{Che10}; \citealp{Zha01}) assume that the Lorentz force follows the temporal profile of the soft X-ray flare that 
often accompanies the CMEs.

In this article we adopt the physical definition for CME aerodynamic drag outlined in \citet{Sac15}, referred
to as Paper 1 from now on, together with a specific model \citep{Kli06} for Lorentz forces to address some of these questions: What is the heliocentric distance range where Lorentz forces dominate? Beyond what heliocentric distance is a 
drag-only model justified? The rest of the article is organized as follows. In Section \ref{S-f} we discuss the forces affecting the CME propagation. 
Section \ref{S-data} provides details of the CME event sample and data obtained from Graduated Cylindrical Shell (GCS) fitting. The analysis and main results are 
outlined in Section \ref{S-res}, followed by discussion and conclusions in Section \ref{S-conc}.

\section{Forces Acting on CMEs} \label{S-f}

Descriptions of CME evolution usually consider an initiation phase comprising the initial CME eruption, which is followed by the propagation phase.
There is an interplay between Lorentz forces, gravity and solar wind aerodynamic drag in the propagation phase; this provides the residual 
acceleration \citep[\textit{e.g.}][]{Zha06,Sub07,Gop13}.  Gravitational forces and plasma pressure are generally
taken to be negligible for flux rope models of CMEs \citep{For00,Ise07}. Lorentz forces are thought to accelerate CMEs up to a few solar radii in the low corona
\citep[\textit{e.g.}][]{Vrs06,Bei11,Car12}, beyond which the solar wind aerodynamic drag takes over. Paper 1
shows that aerodynamic drag accounts for the observed CME trajectory only beyond 15\,--\,50 R$_{\odot}$ for the slow 
(near-Sun speeds $< 900$ km s$^{-1}$) CMEs;
for fast CMEs (near-Sun speed $> 900$ km s$^{-1}$), aerodynamic drag can account for 
their dynamics from 5 R$_{\odot}$ onwards. \citet{Rol16} also 
show that their Drag-based model(DBM) is applicable only beyond a heliocentric distance of 21$\pm$10 R$_{\odot}$.

The forces acting on a CME 
are often represented in the following form (in cgs units):

\begin{eqnarray}
F\,&=& m_{cme} \frac{d^{2}R}{dt^{2}} \nonumber \\
&=&F_{\rm Lorentz} + F_{\rm drag} \nonumber \\
   &=& \biggl\{\biggl[\frac{\pi I^{2}}{c^{2}} \biggl(ln \biggl(\frac{8 R}{b}\biggr)-\frac{3}{2}+ \frac{l_{i}}{2}\biggr)\biggr]\,-\, \frac{(\pi R)I B_{ext}(R)}{c} \biggr\} \nonumber \\
   &&-\,\,\frac{1}{2}\,\, C_{\rm D}\,\, A_{cme} \,\,n_{sw}\,\, m_{p}\, 
 \bigl( V_{cme}-V_{sw} \bigr)\,\,\bigl|V_{cme}-V_{sw}\bigr| \label{eqf}
 \end{eqnarray}
where $F$ is the total force, $m_{\rm cme}$ is the CME mass, $R$ is the heliocentric distance of the leading edge 
of the CME (also often interpreted as the position of the center of mass) and t represents time. $F_{\rm Lorentz}$ is the net Lorentz force acting on the CME in the major radial direction which is given by the term in 
the curly brackets (see \textit{e.g.} \citealp{Sha66}; \citealp{Kli06}). The first term (within the square braces) represents 
the Lorentz self forces ($(1/c)J\times B$, where $J$ is the current density and $B$ the magnetic field) acting on the expanding CME current loop 
\citep[\textit{e.g.}][]{Che89} that accelerate the CME while the second term is
the force due to the external poloidal field ($B_{ext}$) that tends to hold down 
the expanding CME. In the equation, $I$ is the CME current, $c$ is the speed of light, $b$ is CME minor 
radius and $l_{i}$ is the internal inductance. The axial current, $I$, is determined by the conservation of total (\textit{i.e.} 
flux-rope + external) magetic flux.

The term in Equation (\ref{eqf}) involving $C_{\rm D}$ represents the aerodynamic drag experienced by the CME as it propagates through the solar wind. 
The strength of the momentum coupling between the CME and the solar wind is represented by the dimensionless drag 
coefficient, $C_{\rm D}$. We use a non-constant $C_{\rm D}$ given by Equation 7 of Paper 1. For completeness, we include the 
$C_{\rm D}$ definition here: 
\begin{equation}
 C_{\rm D} = 0.148 - 4.3 \times 10^4 \, Re^{-1} + 9.8 \times 10^{-9} \, Re \, ,
\label{eqcdfit}
\end{equation}
where Re is the Reynolds number calculated using the solar wind viscosity expression as described in Paper 1. 
The quantity $A_{\rm cme}$ is the cross sectional area of the CME, $n_{sw}$ is the solar wind density, and $m_{p}$ is the proton mass. 
$V_{\rm cme}$ and $V_{\rm sw}$ denote the CME and solar wind velocities, respectively. Depending on how fast or slow a 
CME is travelling (relative to the solar wind), the solar wind can either ``drag down'' the CME or ``pick it up''. 
Paper 1 finds that fast CMEs (initial velocity $\sim 916$ km s$^{-1}$) are governed primarily by aerodynamic drag from 
as early as $\sim 5.5$ R$_{\odot}$. 
On the other hand, slower CMEs are governed by solar wind
aerodynamic drag only above 15\,--\,50 R$_{\odot}$. 

In this article we analyze a diverse sample of 38 well observed CMEs.
Using measurements with the Graduated Cylindrical Shell (GCS) method for each CME, we determine the heliocentric distance, 
$\widetilde{h}_{0}$, above 
which the CME dynamics is dominated by aerodynamic drag. Using the torus instability (TI) model to describe the Lorentz forces \citep{Kli06},
we address questions such as: Where does the Lorentz 
force peak?, How does the Lorentz force compare with the aerodynamic drag force at and beyond $\widetilde{h}_{0}$? and more.

 
\section{CME Data Sample} \label{S-data}
\subsection{Event Selection}
We investigate CMEs observed during the rising phase of Solar Cycle 24 between 2010 and 2013. The primary data we use is from the 
\textit{Large Angle Spectrometric Coronagaph} \citep[LASCO:][]{Bru95} onboard the \textit{Solar and Heliospheric Observatory} 
mission (SOHO)
and \textit{Sun\,--\,Earth Connection Coronal and Heliospheric Investigation} \citep[SECCHI:][]{How08} coronagraphs and the
\textit{Heliospheric Imagers} (HI) onboard the \textit{Solar TErrestrial RElations Observatory} mission \citep[STEREO;][]{Kai08}. 
\textit{In-situ} measurements are obtained from the \textit{Wind} spacecraft\footnotemark[1], which gives the 
near-Earth parameters for these CMEs. The 38 CMEs we identify for this work have near-Sun speeds 
ranging from 50 km${\rm s}^{-1}$ to 2400 km${\rm s}^{-1}$. Of these, 13 events are partial halo (PH) CMEs and 21 are 
full halo(FH) CMEs as indicated in the SOHO/LASCO CME Catalog\footnotemark[2]. The remaining four events have 
angular width $<120^{\circ}$. All the CMEs in our sample are Earth-directed. 
The respective separations of the STEREO B and A spacecraft from the Earth vary from 71$^\circ$ and 66$^\circ$ in March 2010 to 
about 149$^\circ$ and 150$^\circ$ by the end of 2013.
Along with the LASCO C2 coronagraph, the three point view provides a favourable set up for observing Earth-directed CMEs.
We only include CMEs that have continuous observations in LASCO C2, STEREO-A and -B COR 2, HI1 and HI2. We require that the images from all the instruments must include the CMEs as clear, bright structures.
Events with major distortions and CME\,--\,CME interactions 
were excluded.
\footnotetext[1]{http://omniweb.gsfc.nasa.gov/}
\footnotetext[2]{http://cdaw.gsfc.nasa.gov/CME\_list/}
\subsection{GCS Fitting}

Needless to say, precise information about the 3-dimensional (3D) evolution of CMEs is central to building a good model. 
Early efforts in this direction include those of \citet{Che97} and \citet{Woo99}. The advent of SECCHI/STEREO data facilitated this task 
greatly. 
We use the Graduated Cylindrical Shell (GCS: \citealp {The06}; \citealp{The09}; \citealp{The11}) model to fit the visible CME structure. 
Table \ref{tbl1} lists the GCS fitting parameters for all the CMEs. The serial number of each CME in Table \ref{tbl1}
will be used as a reference to the corresponding event hereafter. The eight events from Paper 1 (marked with an asterisk $\ast$) 
have observations up to the HI2 field of view (FOV), while the remaining events have been fitted up to the HI1 FOV. 
The second and third columns in Table \ref{tbl1} indicate the CME event date and time of the first 
observation in the LASCO C2 FOV. The quantity $h_{0}$ is the height of the leading edge of the CME from the GCS fitting 
technique, at the time of first observation . The CME initial speed $v_{0}$ at $h_{0}$ is calculated by fitting a
third-degree polynomial to the height-time observations. The quantities $n_{wind}$ and $v_{wind}$ are the 
proton number density and solar wind velocity at 1 AU as observed \textit{in-situ} by the \textit{Wind} spacecraft. 
These observed values are extrapolated sunward for use in $F_{\rm drag}$ for calculating $n_{sw}$ and $V_{sw}$ in Equation \ref{eqf}. 
We follow the detailed description given in Paper 1 to calculate of various parameters, such as, $A_{cme}$, $n_{sw}$, $V_{sw}$, and $C_{\rm D}$ 
required in evaluating $F_{\rm drag}$.

GCS parameters like Carrington longitude, $\phi$, and heliographic latitude, $\theta$, along with the tilt, $\gamma$, 
provide details of the position of the source region (SR) and the orientation of the propagating CME. The quantity $\kappa$ is the aspect ratio and $\alpha$ is half 
of the angle between the axes and the legs of the flux-rope. Using the GCS fitted height of the leading edge ($R$), $\kappa$, and $\alpha$ 
at each time instant, other geometrical parameters like CME minor radius, $b$, ratio, $R/b$, elliptical cross-sectional width, and CME area, 
$A_{cme}$, are calculated \citep{The11} to be used in Equation \ref{eqf}. The observed height-time data for each CME in our sample is thus derived from the GCS fitting of images at each time stamp.

Our sample includes 38 CMEs with an initial velocity range $47 <v_{0} < 2400$ km s$^{-1}$. There are 18 CMEs with initial velocities 
$v_{0} > 900$ km s$^{-1}$. 
We call these events ``fast'' and indicate them by a superscript (f) in Tables \ref{tbl1} and \ref{tbl2}.
The fastest event is CME 23 on 27 January 2012, with $v_{0}\sim 2400$ km s$^{-1}$.
The remaning 20 CMEs in the sample having $v_{0}<900$ km s$^{-1}$ are called ``slower'' CMEs.



\section{Analysis and Results} \label{S-res}
Our main aim in this article is to determine the heliocentric distance range(s) where the Lorentz force terms and the aerodynamic drag 
terms (Equation~\ref{eqf}) are respectively dominant.

\subsection{Aerodynamic Drag}

\begin{figure}[h]    
   \centerline{\hspace*{0.065\textwidth}
               \includegraphics[width=0.62\textwidth,clip=]{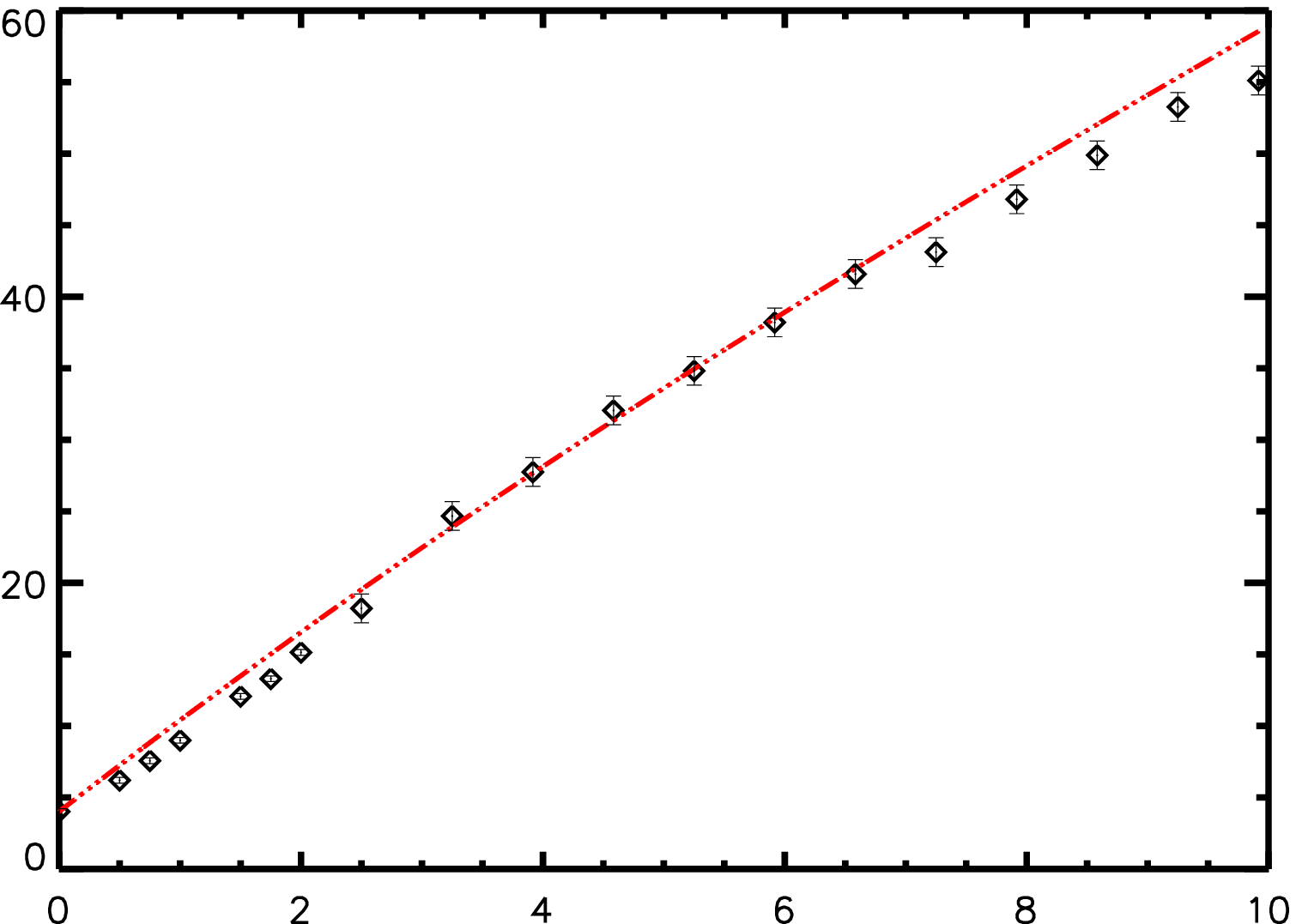}
               \put(-225,60.7){{\rotatebox{90}{{\color{black}\fontsize{8}{8}\fontseries{n}\fontfamily{phv}\selectfont  Height (R$_{\odot}$)}}}}
               \put(-145.8,-6.9){{\rotatebox{0}{{\color{black}\fontsize{8}{8}\fontseries{n}\fontfamily{phv}\selectfont  Elapsed Time (Hours)}}}}
               \hspace*{0.03\textwidth}
               \includegraphics[width=0.63\textwidth,clip=]{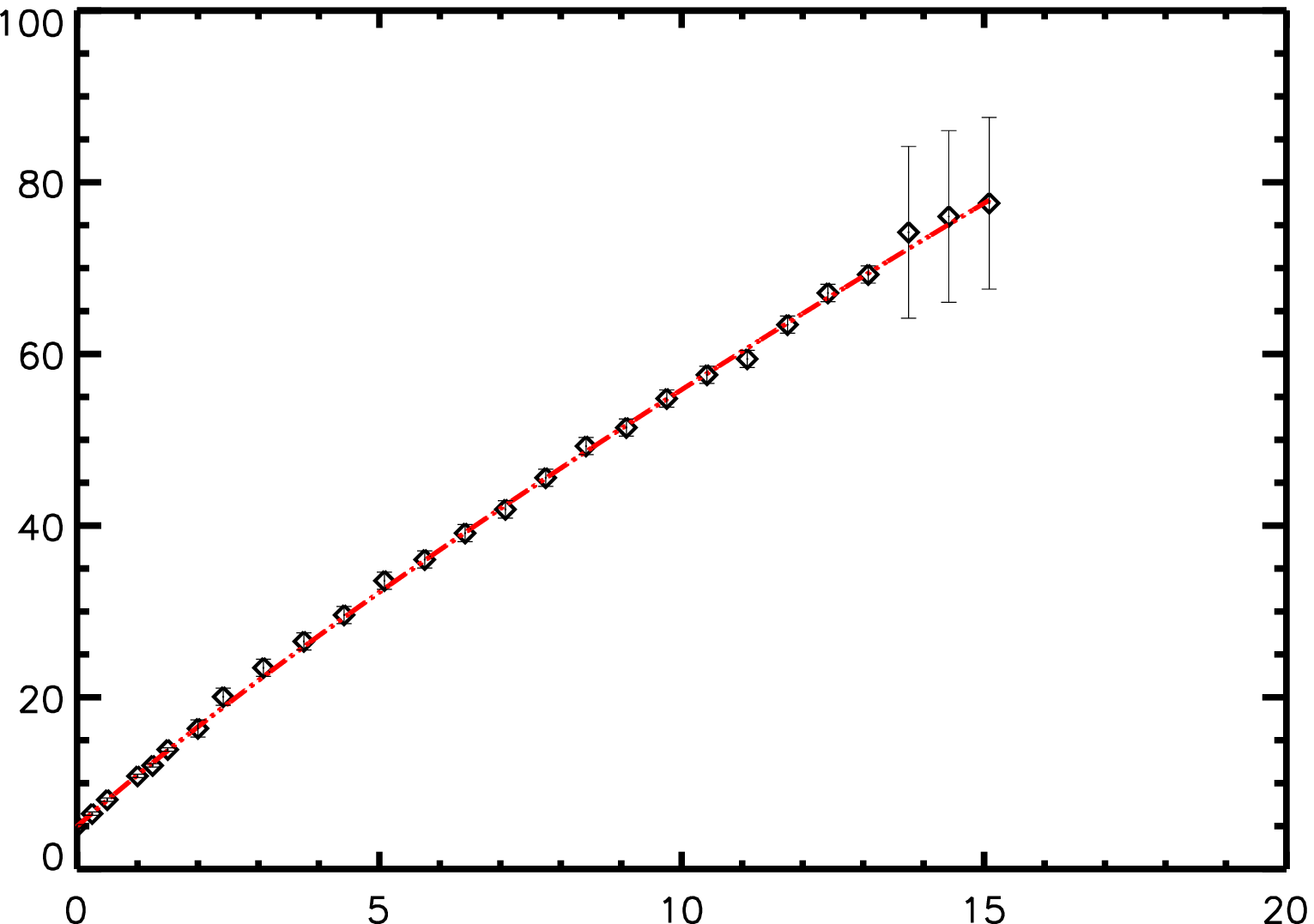}
               \put(-225,60.7){{\rotatebox{90}{{\color{black}\fontsize{8}{8}\fontseries{n}\fontfamily{phv}\selectfont  Height (R$_{\odot}$)}}}}
               \put(-145.8,-6.9){{\rotatebox{0}{{\color{black}\fontsize{8}{8}\fontseries{n}\fontfamily{phv}\selectfont Elapsed Time (Hours)}}}}
              }
           \vspace{-0.49\textwidth}   
               \centerline{\small      
       \hspace{0.15\textwidth} {CME 18}
       \hspace{0.56\textwidth}  {CME 36}
          \hfill}

\vspace{0.461\textwidth}  
\caption{The observed height-time data is shown with diamonds. The red dash-dotted line is the drag model solution when it is initiated from the 
first observed height, $h_{0}$. CME 18 refers to the event on 22 Oct. 2011, with initial speed $\sim$ 1276 km s$^{-1}$ and CME 36 represents an event on 29 
September 2013 with an initial speed $\sim$ 1217 km s$^{-1}$. Both these events are fast CMEs.}
   \label{fig1}
   \end{figure}

\begin{figure} [h]   
   \centerline{\hspace*{0.065\textwidth}
               \includegraphics[width=0.62\textwidth,clip=]{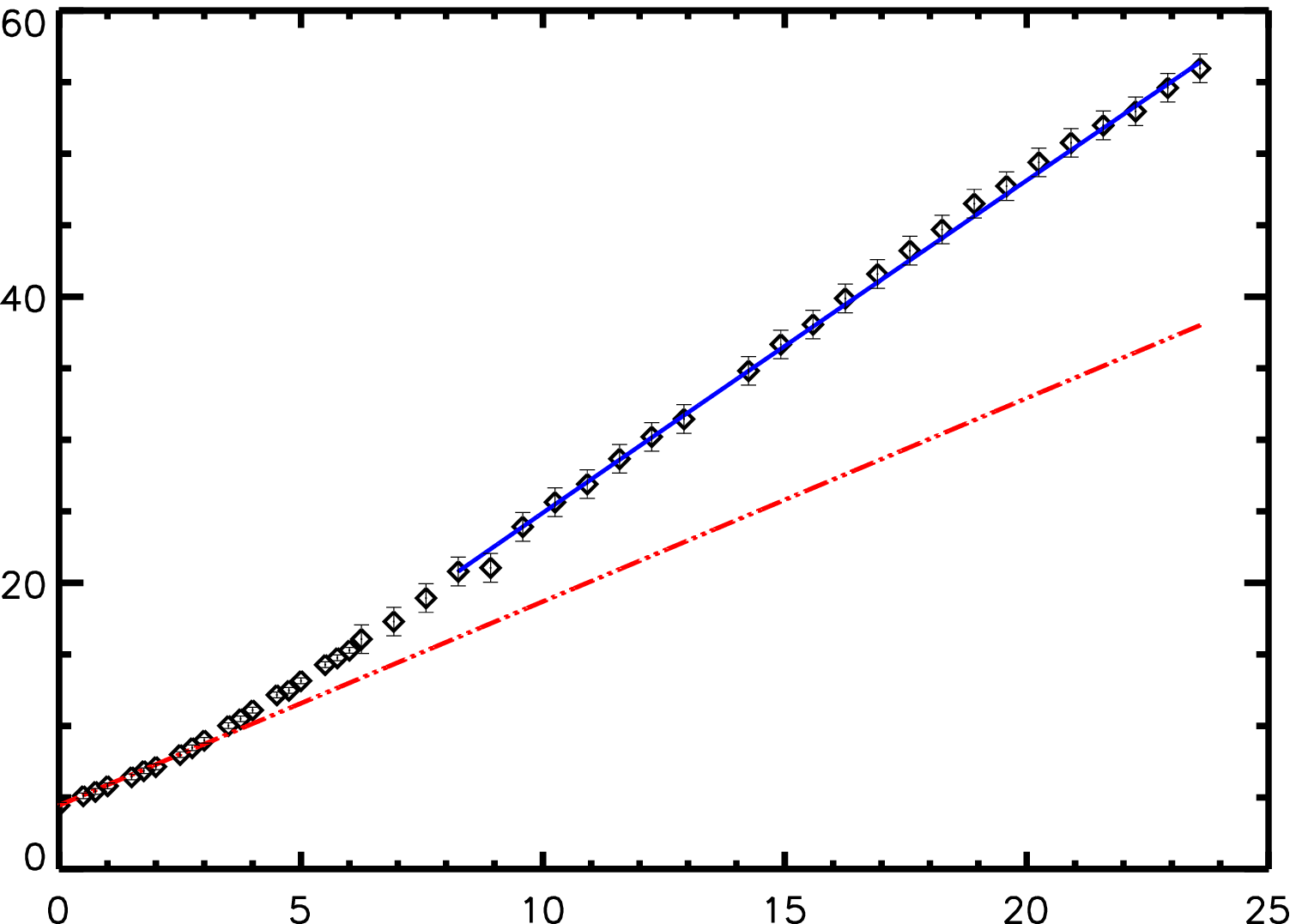}
                 \put(-225,60.7){{\rotatebox{90}{{\color{black}\fontsize{8}{8}\fontseries{n}\fontfamily{phv}\selectfont  Height (R$_{\odot}$)}}}}
               \put(-145.8,-6.9){{\rotatebox{0}{{\color{black}\fontsize{8}{8}\fontseries{n}\fontfamily{phv}\selectfont  Elapsed Time (Hours)}}}}
               \hspace*{0.03\textwidth}
               \includegraphics[width=0.62\textwidth,clip=]{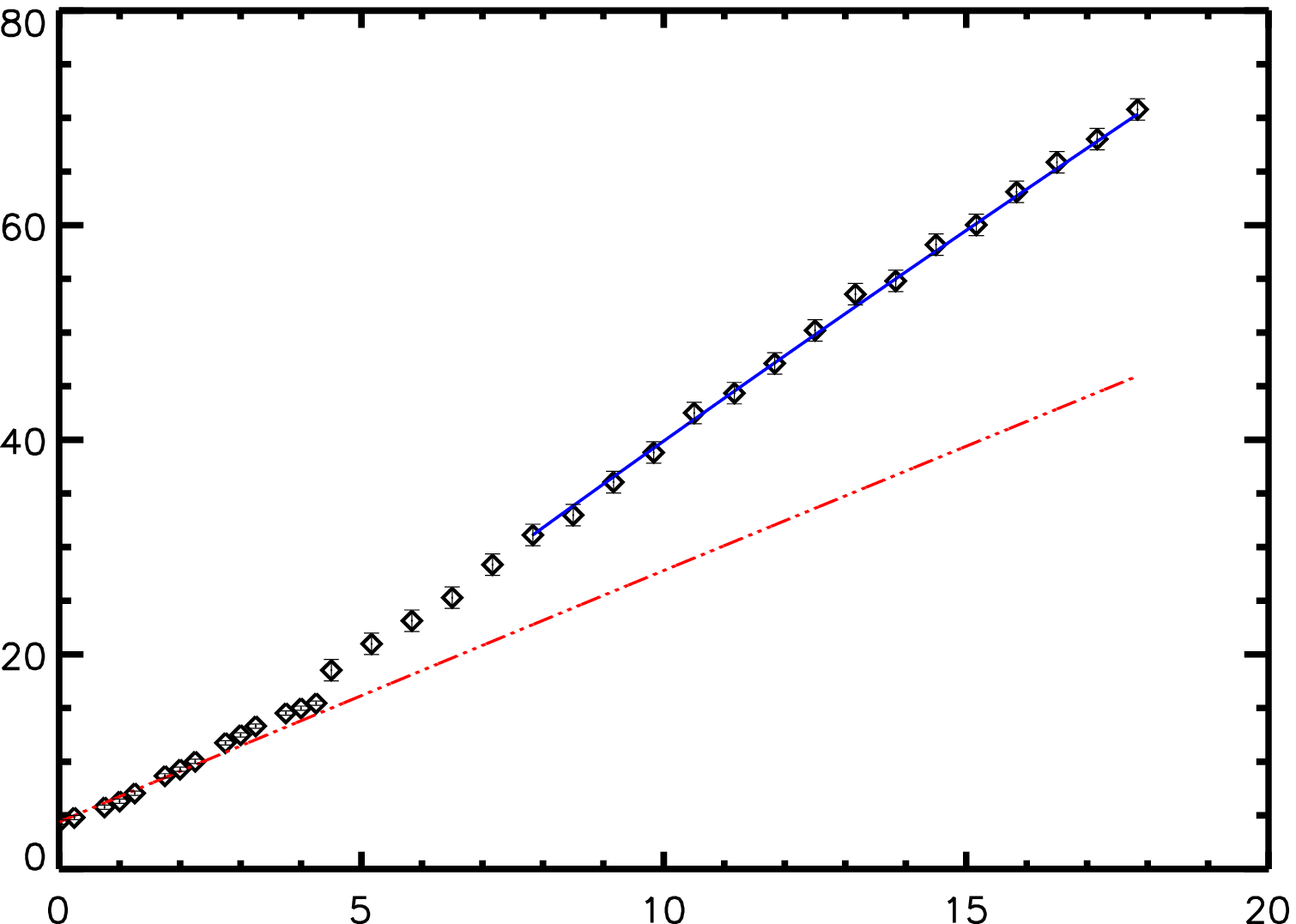}
                \put(-225,60.7){{\rotatebox{90}{{\color{black}\fontsize{8}{8}\fontseries{n}\fontfamily{phv}\selectfont  Height (R$_{\odot}$)}}}}
               \put(-145.8,-6.9){{\rotatebox{0}{{\color{black}\fontsize{8}{8}\fontseries{n}\fontfamily{phv}\selectfont  Elapsed Time (Hours)}}}}              
              }
           \vspace{-0.49\textwidth}   
               \centerline{\small      
       \hspace{0.16\textwidth} {CME 8}
       \hspace{0.56\textwidth}  {CME 29}
          \hfill}
\vspace{0.461\textwidth}  
\caption{The observed height-time data is shown with diamonds. The red dash-dotted line is the drag model solution when it is initiated from the first 
observed height $h_{0}$. The blue solid line shows the
predicted height-time trajectory when the drag model is initiated from height $\widetilde{h}_{0}$. CME 8 refers to the event on 24 January 2011,
with initial speed $\sim$ 276 km s$^{-1}$. CME 29 represents an event on 05 October 2012 with an initial speed 
$\sim$ 461 km s$^{-1}$. Both CMEs 8 and 29 are slow CMEs.}
   \label{fig2}
   \end{figure}

We first try to reconcile the observed CME dynamics with a solar wind aerodynamic drag-only model following the procedure described in 
Paper 1.
In other words, we consider only the $F_{\rm drag}$ term in Equation~\ref{eqf} using 
observationally derived parameters and compare the model solutions with the observed height-time data. 
We find that the drag-only model solutions agree reasonably well with the observed CME 
profile right from the first data point ($h_{0}$) for the fast CMEs (initial velocity $ > 900$ km s$^{-1}$). 
Figure \ref{fig1} shows the height-time plot for CME 18 ($v_{0} \sim 1276$ km s$^{-1}$) and CME 36 
($v_{0} \sim 1217$ km s$^{-1}$), to compare the model results (red dash-dotted line) and data (diamonds).
It is clear that for both these CMEs, solar wind drag explains the observed trajectory quite well from 4 and 4.9 R$_{\odot}$ onwards, 
respectively. This result is representative of all the 18 fast CMEs in our sample. However, this is not true for slower CMEs. 
Figure \ref{fig2} shows the results for two representative slower CMEs (CME 8, $v_{0} \sim 276$ km s$^{-1}$  
and CME 29, $v_{0} \sim 461$ km s$^{-1}$) with the drag-only model initiated from the first observation point.
The disagreement between the data (diamond symbols) and predicted solution
(red dash-dotted line) is obvious, and indicates that the drag-only model, when initiated from the first data point, provides a poor explanation for 
the observed CME dynamics for slower CMEs. As in Paper 1, we then initiate the drag-only model at progressively later heights 
(using observational inputs appropriate to the initiation height). The initiation height at which the drag solution matches the 
observations is denoted by $\widetilde{h}_{0}$ in Table \ref{tbl2}. The model predicted solution (denoted by a solid blue line) shown in Figure \ref{fig2} 
indicates that the drag-only model initiated above $\widetilde{h_{0}}$ ($\sim$ 21 and 31 R$_{\odot}$ for CME 8 and 29 respectively)
provides a good description of the dynamics of these relatively slower CMEs.
We follow this procedure for each event in our sample. The 
quantities $\widetilde{h}_{0}$ and corresponding velocity $\widetilde{v}_{0}$ are listed for each event in Table \ref{tbl2}. 
We use the coefficient of determination (often called $R$ squared) to determine how well the predicted 
model solutions fit the data. Model solutions with $R^{2} > 98\%$ are considered acceptable.
The CME dynamics can be considered to be dominated by solar wind aerodynamic drag above the height $\widetilde{h}_{0}$.
The left panel of Figure \ref{fig33} shows a plot of $\widetilde{h}_{0}$  (Table \ref{tbl2}) \textit{versus} the 
CME initial velocity ($v_{0}$). We see that for CMEs with
$v_{0}<900$ km s$^{-1}$, $\widetilde{h}_{0}$ lies between 12\,--\,50 R$_{\odot}$ while for CMEs with 
$v_{0}>900$ km s$^{-1}$, $\widetilde{h}_{0}$ is same as the initial observed height for the event ($h_{0}$ in Table \ref{tbl1}, which
ranges from 3.9\,--\,8.4 R$_{\odot}$ for the fast CMEs in our sample).
In other words, Figure \ref{fig33} shows that fast CMEs are drag-dominated from 3.9\,--\,8.4 R$_{\odot}$ onwards, while slower CMEs 
are drag dominated only beyond 12\,--\,50 R$_{\odot}$.

\subsection{Lorentz Forces}

If the aerodynamic drag dominates for heliocentric distances $R > \widetilde{h}_{0}$ (\textit{i.e.} it is not necessary to invoke Lorentz 
forces to explain their dynamics), it is natural to investigate the behavior of Lorentz forces for $R < \widetilde{h}_{0}$. 
The first two terms in Equation \ref{eqf} are a feature of most Lorentz force models that deal with CME initiation.
All such models predict that the (total) Lorentz force increases until it 
peaks at a certain heliocentric distance, beyond which it decreases and becomes negligible. Some models tailor the injected poloidal flux (or equivalently, the driving current) so as to achieve 
this Lorentz-force profile \citep{Che10}.
Others, such as the torus instability model \citep{Kli06} rely on the fact that the external Lorentz forces need to decrease 
(with heliocentric distance) faster than a certain rate, in order to ``launch'' the CME. This also results in a Lorentz-force profile that increases
initially and achieves a peak before decreasing. 
\citet{Kli14} have also shown the equivalance of TI and the catastrophe mechanism for CME eruption \citep{For91}.

In this description, the equilibrium position of the flux rope, $h_{eq}$, is defined by a balance between the Lorentz self 
force and the external force. For the sake of concreteness, we adopt $h_{eq} = 1.05$ R$_{\odot}$ in our work. The equilibrium position is 
also defined by an equilibrium current, $I_{eq}$. The current carried by the flux rope at a given $R$ is defined by \citep{Kli06}:

\begin{equation}
  I\,=\,\frac{c^{'}_{eq} I_{eq} h_{eq}}{c^{'} R}\biggl(1+\frac{(c^{'}_{eq}+\frac{1}{2})}{2 c^{'}_{eq}(2-n)}\biggl
  [\biggl(\frac{R}{h_{eq}}\biggr)^{2-n}-1\,\biggr]\biggr) \, ,
\label{eqI}
\end{equation}

where $c^{'}(R)\,=\,\bigl[ln (8R/b)-2+l_{i}/2\bigr]$ and $c^{'}_{eq}\,=\,c^{'}(R=h_{eq})=\bigl[ln (8h_{eq}/b_{eq})-2+l_{i}/2\bigr]$.
The quantity $b$ is the flux rope minor radius. The external (ambient) magnetic field is $\propto R^{-n}$, and $n$ needs to be 
greater than a certain critical value for the torus instability to be operative, causing the flux rope to erupt. The quantity 
$l_{i}$ is the internal inductance of the flux rope, and we use $l_{i}=1/2$.
The equilibrium current, $I_{eq}$, carried by the flux rope is related to the external field $B_{\rm ext}(h_{eq})$ at the equilibirum position via

\begin{equation}
  I_{eq}\,=\,\frac{B_{\rm ext}(h_{eq})h_{eq} c}{c^{'}_{eq}+\frac{1}{2}} \, \nonumber \, .
\end{equation}

For a given value of $n$, the value of $I_{eq}$ (and equivalently $B_{\rm ext}(h_{eq})$) is determined by the 
 condition $F_{\rm drag}(\widetilde{h}_{0})=F_{\rm Lorentz}(\widetilde{h}_{0})$. It constrains the equilibrium current $I_{eq}$ and $n$. 
For a given event, $n$ is chosen to be the minimum value that will ensure that $|F_{\rm drag}|>F_{\rm Lorentz}$ for $R > \widetilde{h}_{0}$. 

Table \ref{tbl2} gives the values of the equilibrium current $I_{eq}$ and $B_{ext}(h_{eq})$ for all CMEs for the corresponding value of 
the decay index n. 
The GCS fits to our observations yield values for the flux rope aspect ratio $R/b$. 
For heliocentric distances below the first observed point $h_{0}$ (which is typically around 3 R$_{\odot}$), we assume 
that $R/b$ is the same as the observed value at the first observed point ($h_{0}$). In other words, we assume that the flux rope 
expands in a self-similar manner from $h_{eq}$ to $h_{0}$; beyond $h_{0}$, we do not have to rely on any such assumption, 
since we have access to the observed values of $R/b$.

%
\subsection{Lorentz Force \textit{versus} Aerodynamic Drag}
As an example, we show a plot of the Lorentz force \textit{versus} heliocentric distance for CMEs 18 and 36 in Figure \ref{fig3} and CMEs 8 and 29 in 
Figure \ref{fig4}. 
The red solid line indicates the points between $h_{eq}$ and $h_{0}$ where we do not have data for $R/b$ 
(in this region we assume that $R/b$ is the same as its value at $h_{0}$) 
and black diamonds indicate points for which we have observationally determined values for $R/b$. Clearly, the Lorentz force
on the flux rope increases from its value at $h_{eq}$ to reach a peak at $h_{peak}$, after which it decreases.
For each CME, the position at which the Lorentz force peaks ($h_{peak}$) is given in Table \ref{tbl2}. The peak is generally between 
1.65 and 2.45 R$_{\odot}$ for the CMEs in our sample.
The green circles in Figures \ref{fig3} and \ref{fig4} indicate the absolute value of solar wind drag force with height 
above $\widetilde{h_{0}}$. The location of $\widetilde{h_{0}}$ is indicated by a blue
dashed vertical line. 
The quantity marked ``$Fall\,\%$'' in Table \ref{tbl2} quantifies the amount by which the Lorentz force at
$\widetilde{h_{0}}$ has fallen from its peak value at $h_{peak}$.
For both the fast CMEs, CME 18 (left panel) and CME 36 (right panel) in Figure \ref{fig3}, the Lorentz force peaks at 
1.95 R$_{\odot}$ with $n=2.1$. The Lorentz force falls by 
35 $\%$ for CME 18 and by 48 $\%$ for CME 36 from $h_{peak}$ up to $\widetilde{h}_{0}$; for these fast CMEs, 
$\widetilde{h}_{0}$ happens to be the same as $h_{0}$. 
For the slower CMEs (CME 8 and CME 29, shown in Figure \ref{fig4}), $n=1.6$ and the Lorentz force peaks at 2.35 R$_{\odot}$ for both 
CMEs. The Lorentz force decreases by as much as 77 $\%$ from its value at $h_{peak}$ by $\widetilde{h}_{0} = 20.8$ R$_{\odot}$, 
beyond which the solar wind drag takes over for CME 8. Similarly, for CME 29, the Lorentz force decreases by 
79 $\%$ from $h_{peak}$ up to $\widetilde{h}_{0} = 31$ R$_{\odot}$.
This is typical of slower CMEs; for all the slower CMEs in Table \ref{tbl2}, the Lorentz force
at $\widetilde{h}_{0}$ (which is 12\,--\,50 R$_{\odot}$) has fallen by around 70-98 $\%$ from its peak value.
While the Lorentz forces peak fairly early on ($h_{peak} \approx$ 1.65\,--\,2.45 R$_{\odot}$) for slow(er) CMEs, this means that they become 
negligible only as far out as 12\,--\,50 R$_{\odot}$.

For all the CMEs in our sample, the right panel of Figure \ref{fig33} shows the percentage 
by which the Lorentz force has fallen at $\widetilde{h}_{0}$ (relative to its peak value) as a function 
of the CME initial speed. Slower CMEs are denoted by blue circles and fast ones by black circles. The $Fall\,\%$ is clearly 
larger for the slower CMEs. Since the fast CMEs are drag dominated from relatively early on 
(left panel of Figure \ref{fig33}), the $Fall\,\%$ is relatively lower.

Reiterating the results summarized in Table \ref{tbl2}.
Column 1 indicates the CME serial number corresponding to the events listed in Table \ref{tbl1}. Column 2 lists the 
height $\widetilde{h}_{0}$ above which the solar wind drag dominates the CME dynamics. For slow CMEs, this height lies in 
the range 12\,--\,50 R$_{\odot}$ while for faster events it is the same as $h_{0}$ in Table \ref{tbl1}.
$\widetilde{v}_{0}$ is the CME speed at height $\widetilde{h}_{0}$. The values of $n$ in column 4 represent the decay index 
for each CME and lie between 1.6 and 3. We note that the fastest CMEs typically have the highest values for $n$. 
The quantity $h_{peak}$ quoted in column 5 gives the position where the Lorentz force peaks; it ranges between 
1.65 and 2.45 R$_{\odot}$. The equilibrium current, $I_{eq}$, in column 6 is in units of 
$10^{10}$ A. The $I_{eq}$ estimates are in agreement with the average axial current calculated by \citet{Sub07}. 
 The quantity $B_{ext}(h_{eq})$ in column 7 is the equilibrium magnetic field at $h_{eq}=1.05$ R$_{\odot}$ in units of G.
$Fall\,\%$ in column 8 describes the amount by which the Lorentz force at $\widetilde{h}_{0}$ has decreased relative to 
its peak value. For slow CMEs, the percentage fall is between 70\,--\,98 $\%$ while 
for faster CMEs, it is between 20\,--\,60 $\%$. The last column indicates the quantity 
$F_{\rm diff} \,=\,((F_{\rm drag}-F_{\rm Lorentz})/F_{\rm drag})\times 100 \%$ evaluated at 40 R$_{\odot}$ (except for CME 11, where it is evaluated
at 50 R$_{\odot}$).

\section{Discussion and Conclusions} \label{S-conc}
\subsection{Discussion}

Our main aim in this paper is to quantify the relative contributions of Lorentz forces and solar wind aerodynamic drag on 
CMEs as a function of heliocentric distance. Since these are the two main forces thought to be responsible 
for CME dynamics, it is essential to know their relative importance to build reliable models for CME Earth arrival 
time and speed. It is known that aerodynamic drag dominates CME dynamics only beyond distances as large as 15\,--\,50 R$_{\odot}$ 
for all but the fastest CMEs (Paper 1). This trend has also been confirmed by independent studies using an 
empirical fitting parameter for aerodynamic drag (\textit{e.g.} Temmer et al 2015). One would assume that Lorentz forces 
are dominant below these heliocentric distances (and negligible above it), but this has not been explicitly confirmed 
so far. To the best of our knowledge, this is the first systematic study in this regard using a diverse CME sample.

We use a sample of 38 CMEs that are well observed by the SECCHI coronagraphs and 
the heliospheric imagers onboard STEREO and the LASCO coronagraphs onboard SOHO. We use detailed geometrical parameters from GCS 
fitting to the CMEs. Our prototypical models for aerodynamic drag and Lorentz forces are shown in Equation \ref{eqf}. The model 
for aerodynamic drag follows the physical definition outlined in Paper 1, and the model for Lorentz forces follows the TI model
of \citet{Kli06}.  
Using only the aerodynamic drag term, we compute the heliocentric distance $\widetilde{h}_{0}$ beyond which solar 
wind drag can be considered to be the only force influencing the CME dynamics. This calculation makes use of several observational 
inputs for each CME: the ambient solar wind density and velocity, GCS fitted height, velocity and area for each CME. Subsequently, 
we use only the Lorentz force term. Using observational data for the aspect ratio of the CME flux rope, we determine the 
heliocentric distance $h_{peak}$ at which the Lorentz force attains its peak value. We also determine the percentage by
which the Lorentz force decreases from its peak value at $h_{peak}$ up to $\widetilde{h}_{0}$ (beyond which aerodynamic drag becomes 
the dominant force). Table \ref{tbl2} summarizes all our results. 

\begin{figure}[h]    
   \centerline{\hspace*{0.065\textwidth}
               \includegraphics[width=0.61\textwidth,clip=]{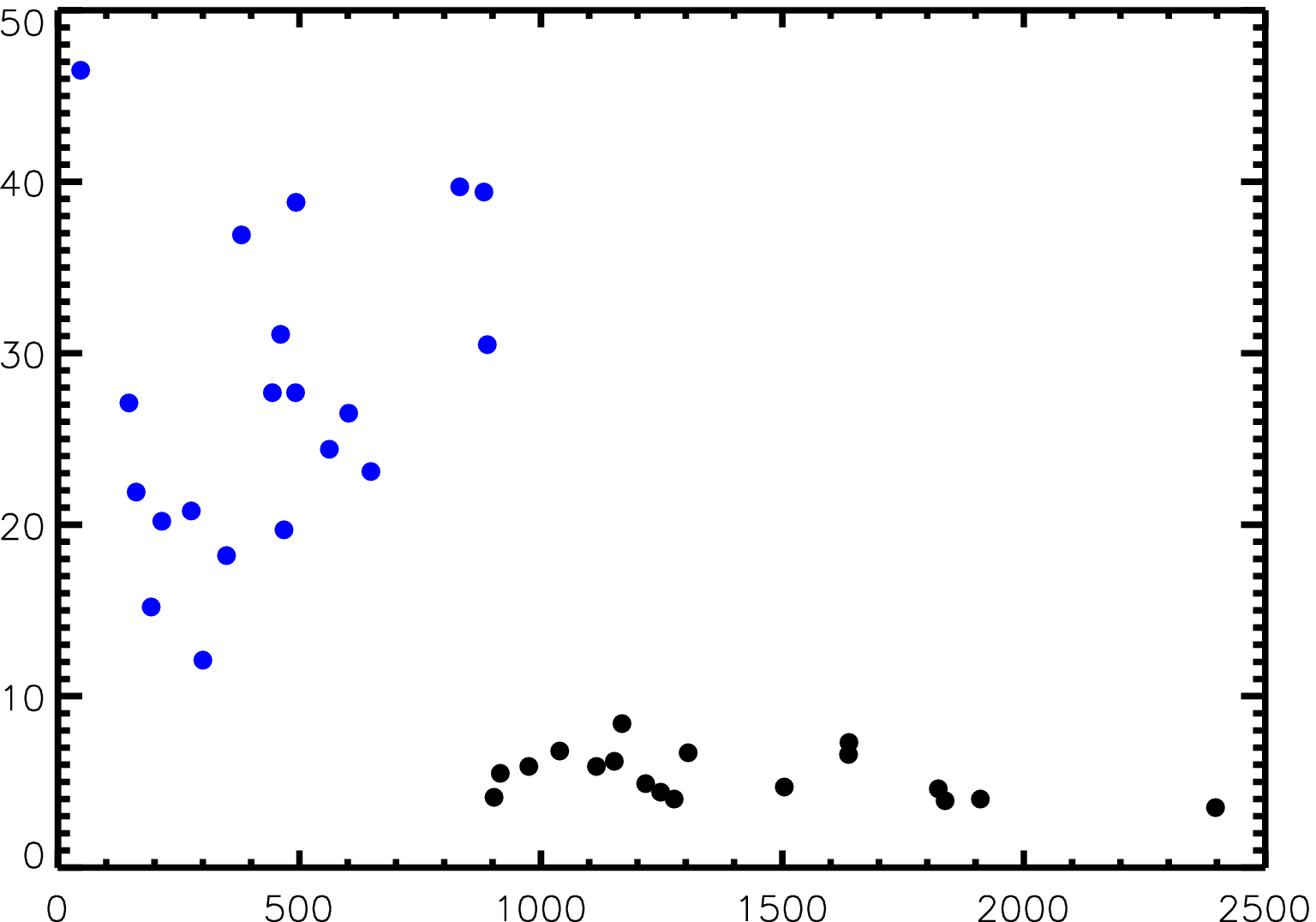}
               \hspace*{0.01\textwidth}
               \put(-230,65.7){{\rotatebox{90}{{\color{black}\fontsize{9}{9}\fontseries{n}\fontfamily{phv}\selectfont  $\widetilde{h}_{0}$ (R$_{\odot}$)}}}}
               \put(-125.8,-7.9){{\rotatebox{0}{{\color{black}\fontsize{9}{9}\fontseries{n}\fontfamily{phv}\selectfont  $v_{0}$ (km s$^{-1})$}}}}
               \includegraphics[width=0.62\textwidth,clip=]{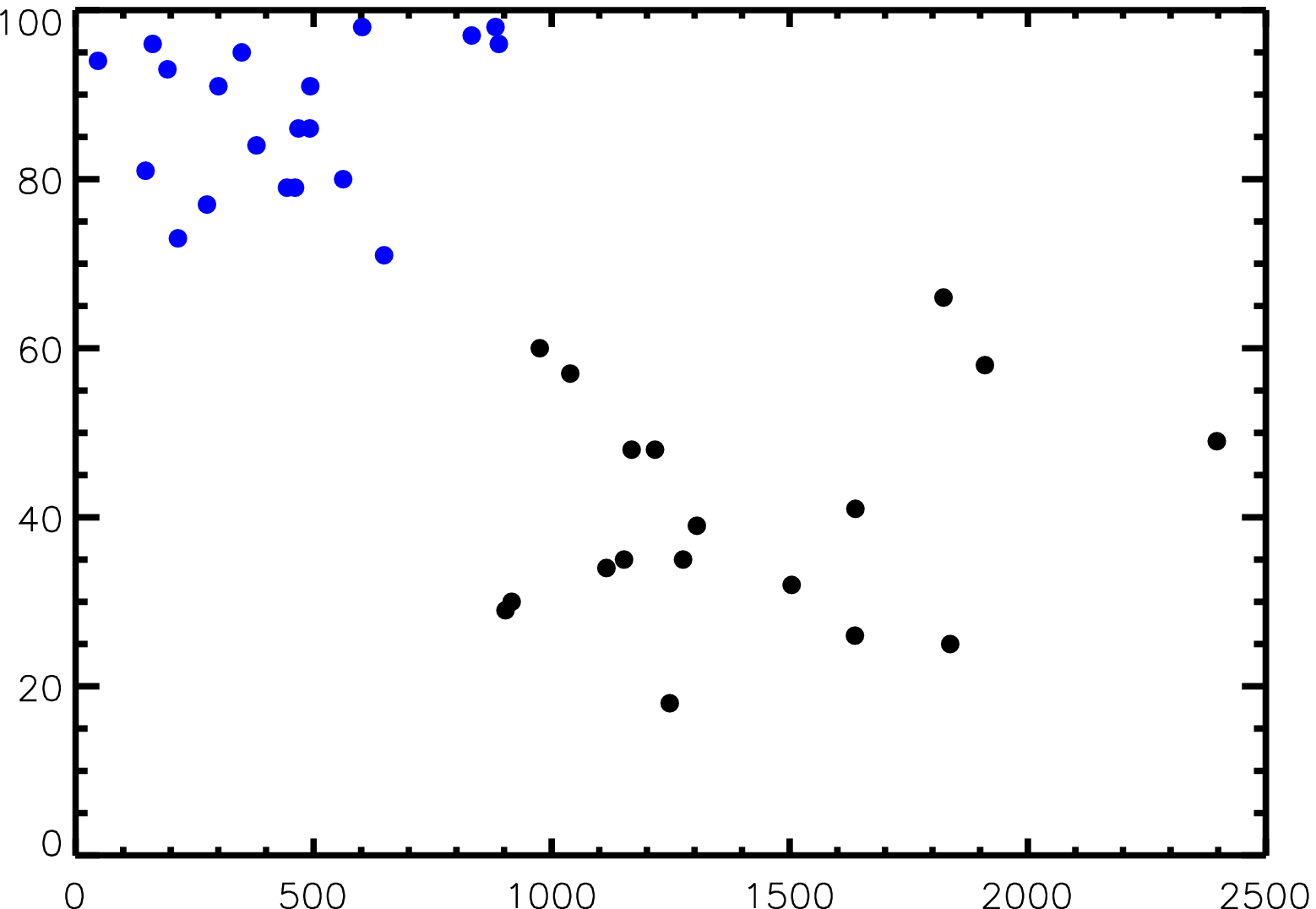}
               \put(-220,65.7){{\rotatebox{90}{{\color{black}\fontsize{7.7}{9}\fontseries{n}\fontfamily{phv}\selectfont   $Fall\,\%$}}}}
               \put(-125.8,-7.9){{\rotatebox{0}{{\color{black}\fontsize{9}{9}\fontseries{n}\fontfamily{phv}\selectfont  $v_{0}$ (km s$^{-1})$}}}}
              }
           \vspace{-0.465\textwidth}   
               \centerline{\small      
           \hfill}

\vspace{0.442\textwidth}  
\caption{Plot of initiation height ($\widetilde{h}_{0}$) and $Fall\,\%$ \textit{versus} the CME initial speed. Left panel shows the quantity 
$\widetilde{h}_{0}$ (from where solar wind drag dominates) as a function of CME initial velocity, $v_{0}$. The right panel depicts the percentage fall 
in the Lorentz force from its peak value to its value at $\widetilde{h}_{0}$ (\%) as a function of CME initial velocity, $v_{0}$. 
Symbols in blue represent slow CMEs (\textit{i.e.} $v_{0}<900$ km s$^{-1}$) and symbols in black represent fast CMEs ($v_{0}>900$ km s$^{-1}$).
See Table \ref{tbl2} for the values in the figure.}

   \label{fig33}
   \end{figure}

\begin{figure}[h]    
   \centerline{\hspace*{0.065\textwidth}
               \includegraphics[width=0.61\textwidth,clip=]{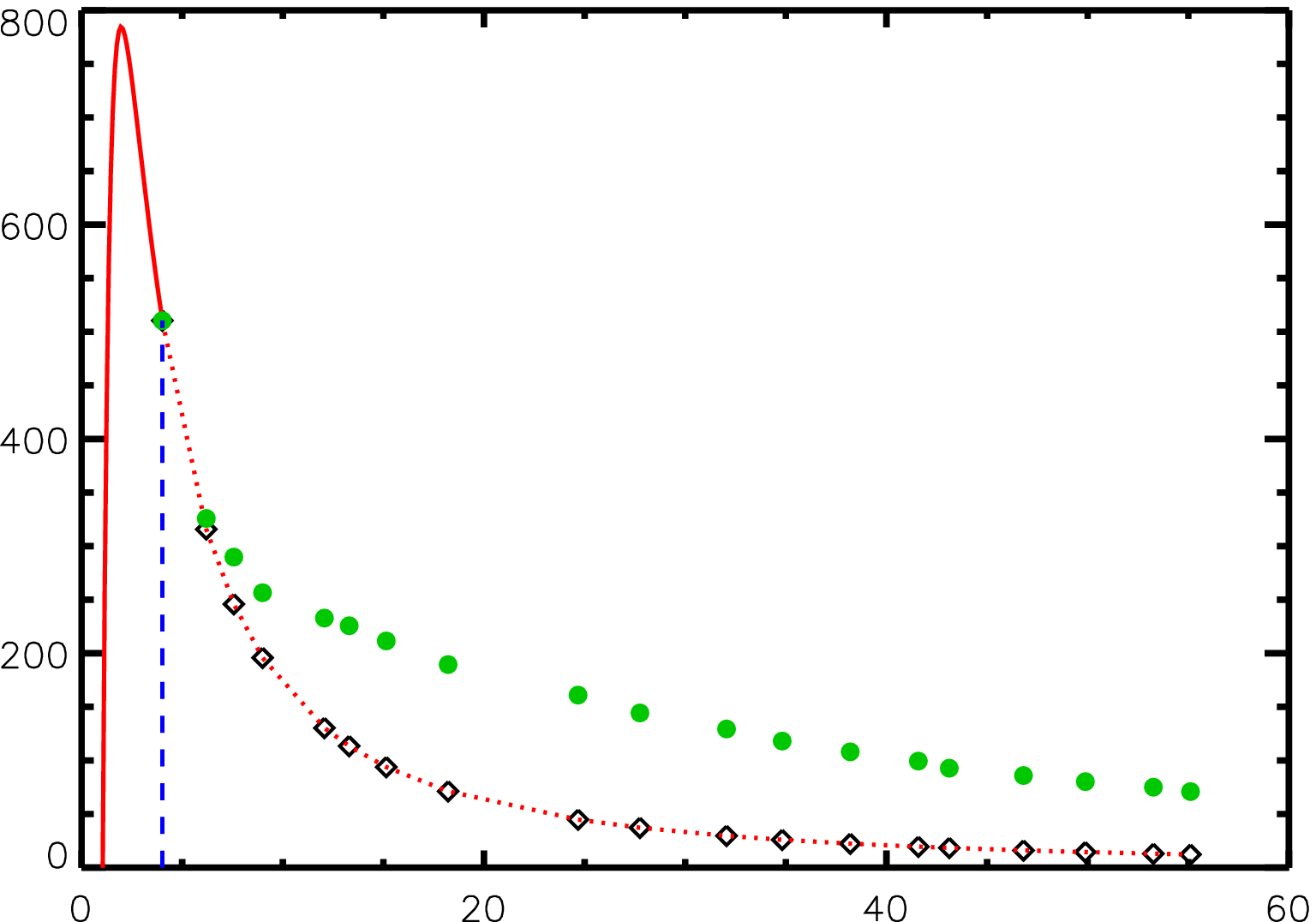}
                \put(-222,40.7){{\rotatebox{90}{{\color{black}\fontsize{8}{7.}\fontseries{n}\fontfamily{phv}\selectfont  Force (10$^{17}$ dyn)}}}}
               \put(-155.8,-8.2){{\rotatebox{0}{{\color{black}\fontsize{9}{9}\fontseries{n}\fontfamily{phv}\selectfont  Heliocentric distance ($R$) (R$_{\odot}$)}}}}
               \hspace*{0.03\textwidth}
               \includegraphics[width=0.61\textwidth,clip=]{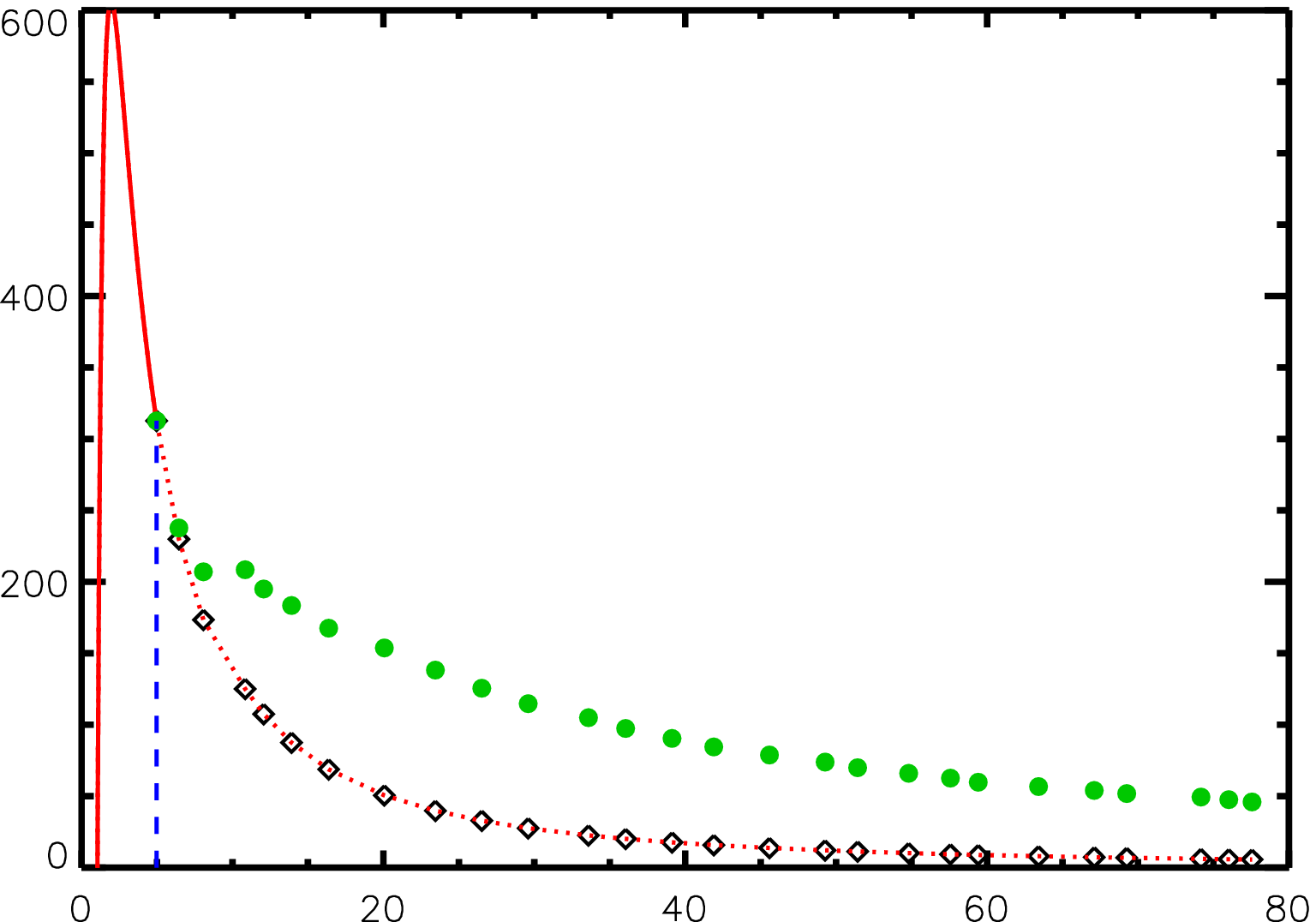}
               \put(-222,40.7){{\rotatebox{90}{{\color{black}\fontsize{8}{7.}\fontseries{n}\fontfamily{phv}\selectfont  Force (10$^{17}$ dyn)}}}}
               \put(-155.8,-8.2){{\rotatebox{0}{{\color{black}\fontsize{9}{9}\fontseries{n}\fontfamily{phv}\selectfont  Heliocentric distance ($R$) (R$_{\odot}$)}}}}              
              }
           \vspace{-0.482\textwidth}   
               \centerline{\small      
       \hspace{0.16\textwidth} {CME 18}
       \hspace{0.56\textwidth}  {CME 36}
          \hfill}

\vspace{0.454\textwidth}  
\caption{Comparison of Lorentz and drag forces for fast CMEs. 
The open diamond symbols represent the Lorentz force values derived observationally starting from $h_{0}$. The red solid line indicates
the Lorentz force values for heights between $h_{eq}$ and $h_{0}$. The filled green circles represent the absolute value of the solar wind drag force. The dashed vertical line (blue) indicates the height $\widetilde{h}_{0}$ at which the solar 
wind drag force takes over. This height is the first observation point for both CMEs 18 and 36. For CME 18 $\widetilde{h}_{0} \sim 4$ R$_{\odot}$,
while for CME 36 $\widetilde{h}_{0} \sim  4.9$ R$_{\odot}$. }
   \label{fig3}
   \end{figure}

\begin{figure}[h]    
   \centerline{\hspace*{0.065\textwidth}
               \includegraphics[width=0.618\textwidth,clip=]{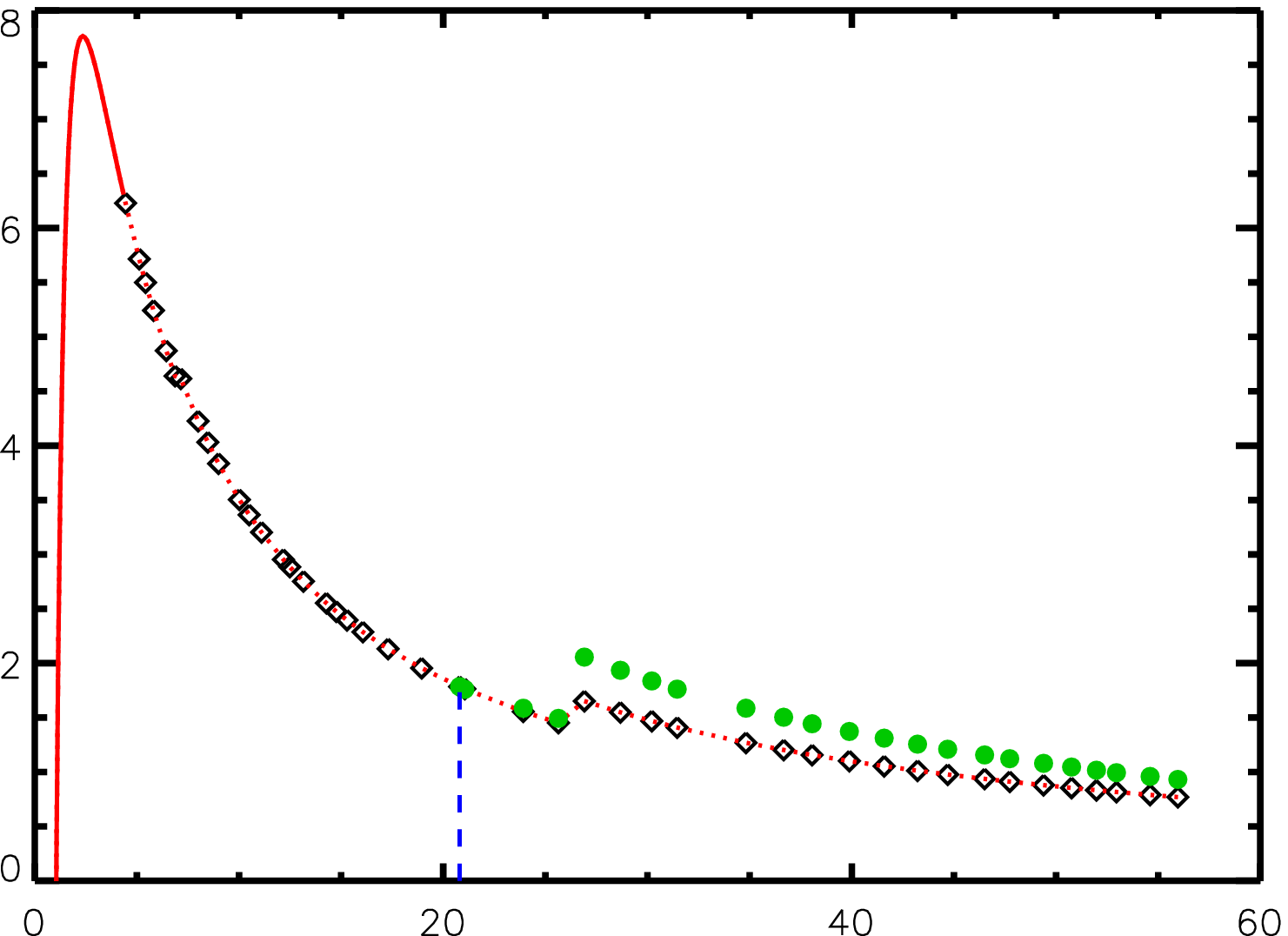}
                 \put(-222,40.7){{\rotatebox{90}{{\color{black}\fontsize{8}{7.}\fontseries{n}\fontfamily{phv}\selectfont  Force (10$^{17}$ dyn)}}}}
               \put(-155.8,-8.2){{\rotatebox{0}{{\color{black}\fontsize{9}{9}\fontseries{n}\fontfamily{phv}\selectfont  Heliocentric distance ($R$)(R$_{\odot}$)}}}}
               \hspace*{0.03\textwidth}
               \includegraphics[width=0.64\textwidth,clip=]{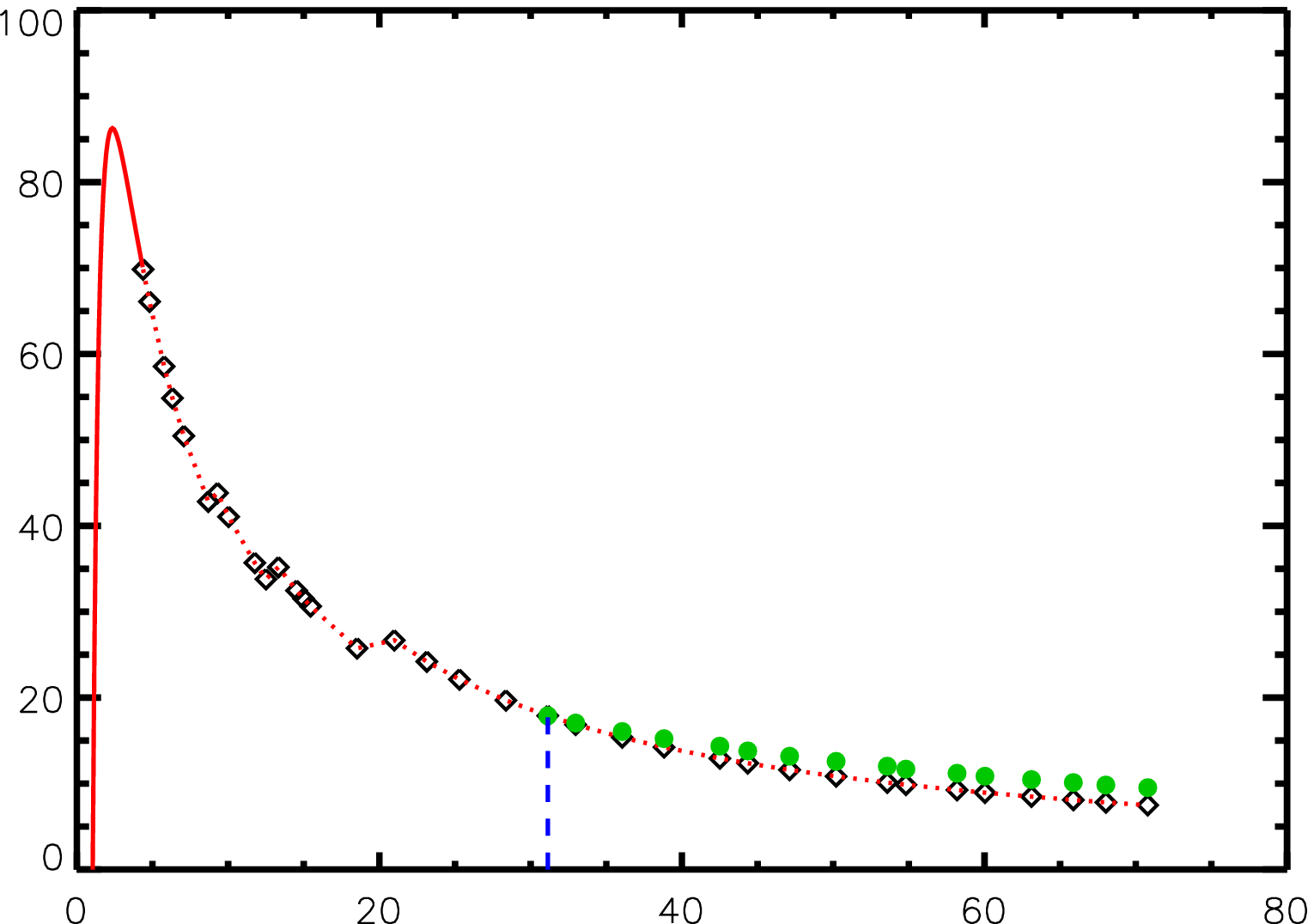}
                \put(-228,40.7){{\rotatebox{90}{{\color{black}\fontsize{8}{7.}\fontseries{n}\fontfamily{phv}\selectfont  Force (10$^{17}$ dyn)}}}}
               \put(-155.8,-8.2){{\rotatebox{0}{{\color{black}\fontsize{9}{9}\fontseries{n}\fontfamily{phv}\selectfont  Heliocentric distance ($R$)(R$_{\odot}$)}}}}   
              }
           \vspace{-0.502\textwidth}   
               \centerline{\small      
       \hspace{0.14\textwidth} {CME 8}
       \hspace{0.56\textwidth}  {CME 29}
          \hfill}

\vspace{0.48\textwidth}  
\caption{Comparison of Lorentz and drag forces for the slow CMEs (8 and 29). 
The symbols and linestyles are the same as in Figure~\ref{fig3}.}
   \label{fig4}
   \end{figure}

\begin{figure}[h]    
\begin{center}
 
               \includegraphics[width=1.\textwidth,clip=]{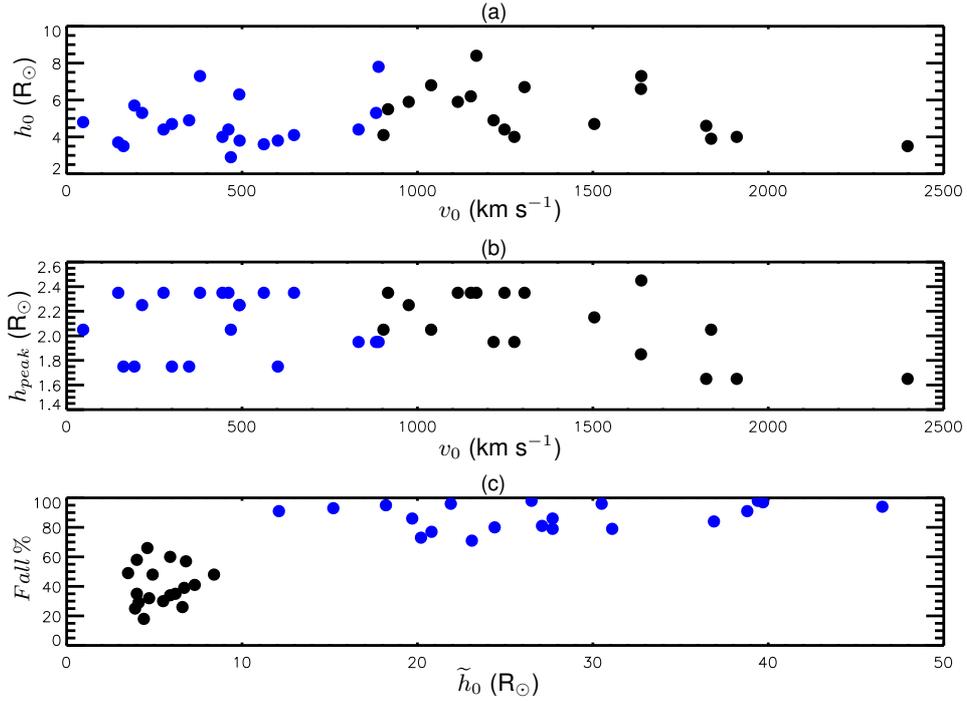}
               \put(-180,245.7){{\rotatebox{0}{{\color{black}\fontsize{8}{8}\fontseries{n}\fontfamily{phv}\selectfont  (a)}}}}
               \put(-355,200.7){{\rotatebox{90}{{\color{black}\fontsize{9.0}{9}\fontseries{n}\fontfamily{phv}\selectfont  $h_{0}$ (R$_{\odot}$)}}}}
               \put(-195.8,170.9){{\rotatebox{0}{{\color{black}\fontsize{9}{9}\fontseries{n}\fontfamily{phv}\selectfont  $v_{0}$ (km s$^{-1}$)}}}}
               \put(-180,156.7){{\rotatebox{0}{{\color{black}\fontsize{8}{8}\fontseries{n}\fontfamily{phv}\selectfont  (b)}}}}
               \put(-357,100.7){{\rotatebox{90}{{\color{black}\fontsize{9.0}{9}\fontseries{n}\fontfamily{phv}\selectfont  $h_{peak}$ (R$_{\odot}$)}}}}
               \put(-195.8,80.9){{\rotatebox{0}{{\color{black}\fontsize{9}{9}\fontseries{n}\fontfamily{phv}\selectfont  $v_{0}$ (km s$^{-1}$)}}}}
               \put(-180,67.7){{\rotatebox{0}{{\color{black}\fontsize{8}{8}\fontseries{n}\fontfamily{phv}\selectfont  (c)}}}}
               \put(-355,25.7){{\rotatebox{90}{{\color{black}\fontsize{7.7}{9}\fontseries{n}\fontfamily{phv}\selectfont  $Fall\,\%$}}}}
               \put(-188.8,-8.){{\rotatebox{0}{{\color{black}\fontsize{9}{9}\fontseries{n}\fontfamily{phv}\selectfont  $\widetilde{h}_{0}$ (R$_{\odot}$)}}}}
               \end{center}
               \vspace{-0.5cm}
\caption{Summary of some of the results in Table \ref{tbl2}. The blue circles represent quantities for slow CMEs and black circles represent fast CMEs.
Panels a and b  
plot the quantities $h_{0}$ and $h_{peak}$ respectively as a function of CME initial velocity, $v_{0}$. 
Panel c shows the percentage decrease, $Fall\,\%$ in Lorentz force (between its peak and $\widetilde{h}_{0}$) as a function 
of  $\widetilde{h_{0}}$ for both slow (blue circles) and fast 
(black circles) CMEs. }
   \label{fig7}
   \end{figure}

\begin{figure}[h]    
   \centerline{\hspace*{0.065\textwidth}
                               \includegraphics[width=0.69\textwidth,clip=]{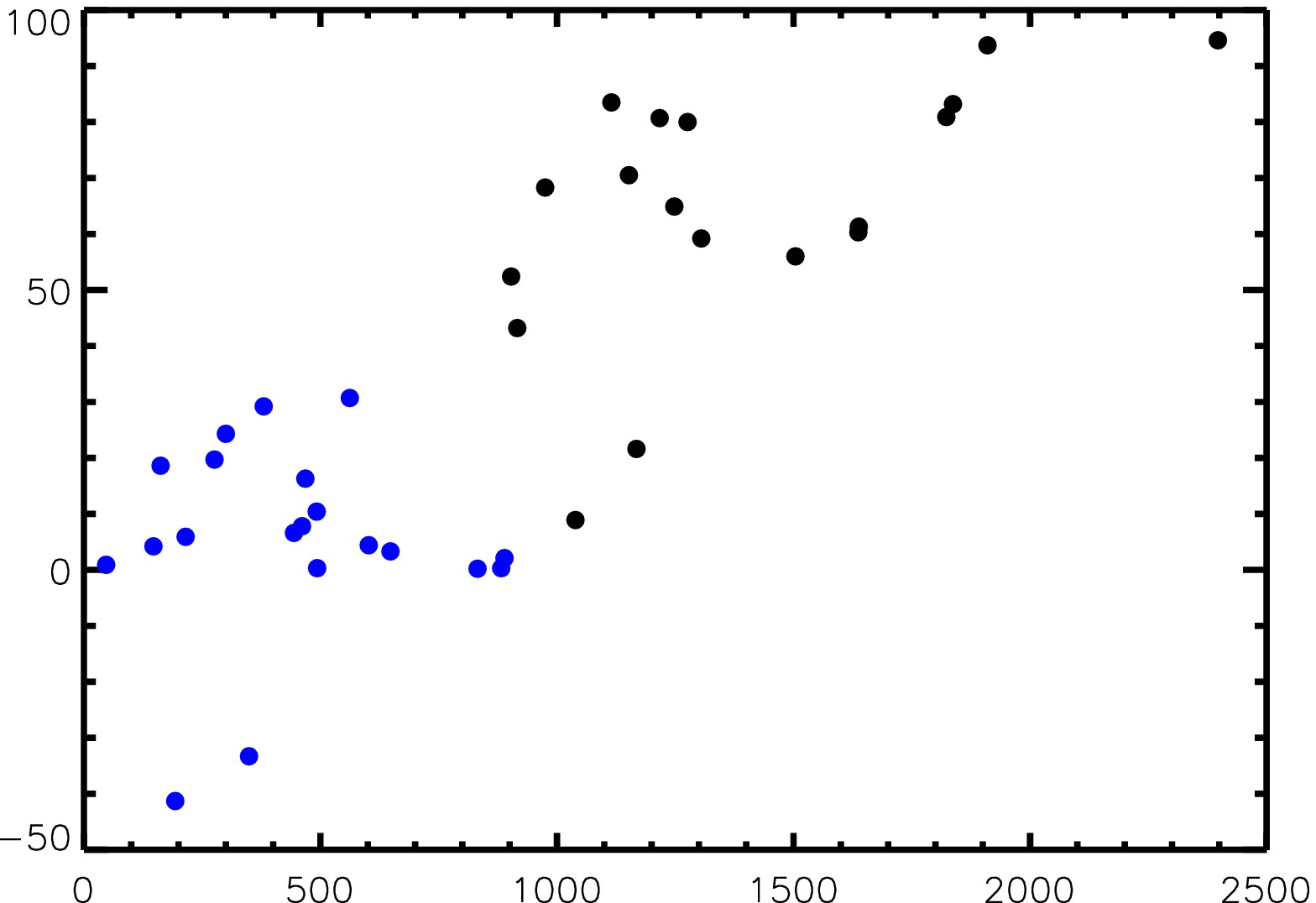}
               \put(-255,27.7){{\rotatebox{90}{{\color{black}\fontsize{8.}{8.}\fontseries{n}\fontfamily{phv}\selectfont 
               $F_{\rm diff}\,=\,100 \times \frac{(F_{\rm drag}-F_{\rm Lorentz})}{F_{\rm drag}} \%$}}}}
                \put(-140.8,-10.9){{\rotatebox{0}{{\color{black}\fontsize{9.}{9}\fontseries{n}\fontfamily{phv}\selectfont  $v_{0}$ (km s$^{-1}$)}}}}
         \hspace*{0.03\textwidth}
               \includegraphics[width=0.7\textwidth,clip=]{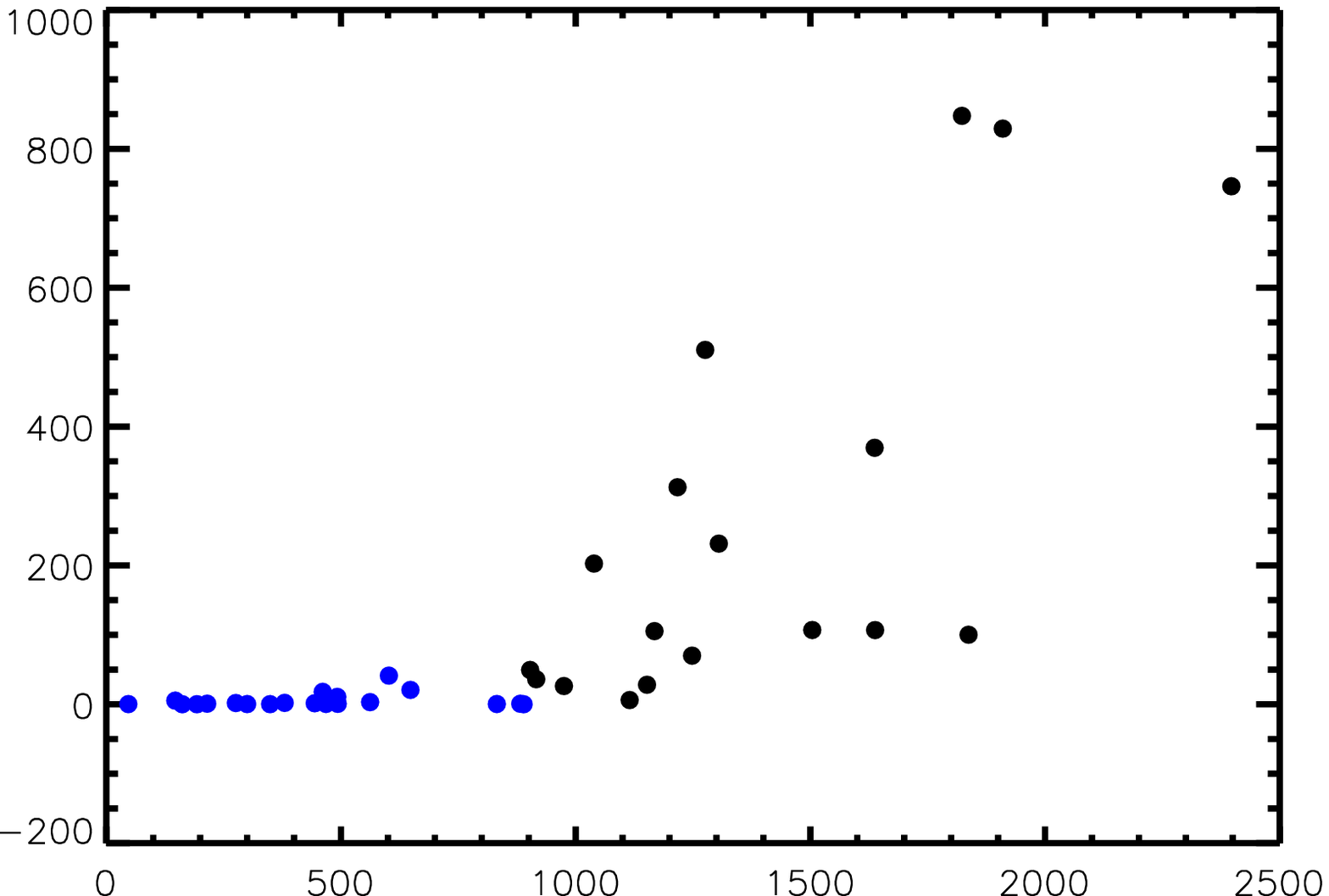}
                \put(-248,45.7){{\rotatebox{90}{{\color{black}\fontsize{8}{8.}\fontseries{n}\fontfamily{phv}\selectfont  Force (10$^{17}$ dyn)}}}}
   \put(-138.8,-10.9){{\rotatebox{0}{{\color{black}\fontsize{9.}{9}\fontseries{n}\fontfamily{phv}\selectfont  $v_{0}$ (km s$^{-1}$)}}}}
              }
           \vspace{-0.545\textwidth}   
               \centerline{\small      
       \hspace{0.14\textwidth} {{\bf (a)}}
       \hspace{0.69\textwidth}  {{\bf (b)}}
          \hfill}

\vspace{0.51\textwidth}  
\caption{Plot of the relative difference between the two forces at $\widetilde{h}_{0}$ and variation of the drag force 
magnitude \textit{versus} CME initial velocity. Panel a shows the percentage difference between drag and Lorentz forces as a function of CME initial 
velocity, $v_{0}$. $F_{\rm diff}$ is calculated at 40 R$_{\odot}$ for all CMEs except CME 11, for which it is evaluated at 50 R$_{\odot}$. 
Panel b shows the absolute value of drag force at $\widetilde{h}_{0}$ for all CMEs with respect to the CME initial velocity, $v_{0}$. 
Details are 
described in the text. The blue circles represent slow CMEs and the black ones represent fast CMEs. }
   \label{fig8}
   \end{figure}

Some of the trends revealed by our results are depicted graphically in Figures \ref{fig33}, \ref{fig7}, and \ref{fig8}. 
Blue circles indicate slow CMEs while the black circles represent fast events. As discussed earlier, 
Figure~\ref{fig33} shows that aerodynamic drag dominates the dynamics of fast CMEs from a few R$_{\odot}$ onwards,
whereas it is dominant for slow CMEs only beyond 12\,--\,50 R$_{\odot}$. 

Figure \ref{fig7}a shows the first observed
height, $h_{0}$, of each CME as a function of its initial speed $v_{0}$. There does not seem to be a definite 
distinction between fast and slow CMEs in this regard. It is possible, however, that the limited 
time cadence of the LASCO C2 coronagraphs affects the values of $h_{0}$ for fast CMEs. We note that the first observed height for about 60 $\%$ of the CMEs lies 
between 2.9 and 5.0 R$_{\odot}$.

As depicted in Figures \ref{fig3} and \ref{fig4}, the Lorentz force profile shows a 
steep increase from $h_{eq}$ until it peaks at $h_{peak}$, beyond which it decreases.
Figure \ref{fig7}b is a scatterplot of the position of 
the Lorenz force peak ($h_{peak}$, see Table \ref{tbl2}) as a function of the CME initial speed, $v_{0}$. 
The value of  $h_{peak}$ is between 1.65 and 2.45 R$_{\odot}$ for all CMEs, with no noticeable 
trend distinguishing slow and fast ones.

As shown in the right panel of Figure~\ref{fig33}, the percentage decrease, $Fall\, \%$, of the Lorentz force 
at $\widetilde{h}_{0}$ (relative to its value at $h_{peak}$) is considerably higher for 
slow CMEs than it is for fast ones. Figure \ref{fig7}c shows a different way of 
visualising this data - the quantity $Fall\,\%$ is plotted as a function of $\widetilde{h}_{0}$. 
It shows that the percentage decrease is larger for CMEs with larger $\widetilde{h}_{0}$ (the slow ones) 
than it is for those with relatively smaller values of $\widetilde{h}_{0}$ (the fast ones).

The drag-only model accounts well for the CME trajectory when initiated at $\widetilde{h}_{0}$ (or beyond).
This implies that other forces (such as Lorentz forces) are not important beyond this height. We find this 
to be true for 36 of the 38 CMEs in our sample. CMEs 4 and 10 (which are slow) are the only exceptions. 
However, we find that the difference between the drag force and the Lorentz force beyond $\widetilde{h}_{0}$ is much more 
pronounced for fast CMEs than for slow ones (\textit{e.g.} Figures \ref{fig3} and \ref{fig4}). In order to quantify this, 
we compute the quantity $F_{\rm diff}\,= \,100 \times (F_{\rm drag} - F_{\rm Lorentz})/F_{\rm drag}$ for all the CMEs in our list.
This quantity is plotted in Panel a of Figure~\ref{fig8} as a function of the 
CME initial velocity $v_{0}$; as before, blue circles represent slow CMEs while black ones represent fast ones.
We show the relative percentage difference for all events at 40 R$_{\odot}$ except for CME 11. Since $\widetilde{h}_{0} \sim
46$ R$_{\odot}$ for CME 11, $F_{\rm diff}$ is evaluated at 50 R$_{\odot}$ for this event. The drag force is 50\,--\,90\% larger at 40 R$_{\odot}$ than the Lorentz force 
for most of the fast events. This justifies the success of the drag-only model for fast events. On the other hand, 
this number ranges from 0.2\% and 30\% for the slower events. Evidently, 
for some of the slower CMEs, the computed Lorentz force is only slightly smaller than the drag force, even well beyond 
$\widetilde{h}_{0}$. For the slow CMEs 4 and 10, the Lorentz force is 
in fact larger than the solar wind drag force in magnitude. 
Figure \ref{fig8}b shows a plot of the absolute magnitude of the solar wind drag force (in units of $10^{17}$ dyn) 
for all CMEs at $\widetilde{h}_{0}$ \textit{versus} the CME initial velocity, $v_{0}$.
Since a drag-only model describes the data well for all the events in our 
list, it follows that the Lorentz force we compute for some of the slower CMEs is an over-estimate. Our Lorentz force 
computations assume that the total magnetic flux is frozen in. The time evolution of the enclosed current $I$, follows from this assumption. 
However, this assumption might not be accurate -  some of the magnetic energy might be expended in CME expansion and/or heating of the CME plasma.
Our results suggest that such effects might be especially important for slower CMEs.

\subsection{Conclusions}
Our main conclusions are:
\begin{itemize}
\item
Our analysis shows that a model that includes only the solar wind aerodynamic drag accurately describes the trajectory of fast CMEs from very early on. 
The converse is true for slower CMEs. The distance $\widetilde{h}_{0}$ beyond which the solar wind aerodynamic drag dominates over Lorentz forces can be as small as 
3.5\,--\,4 R$_{\odot}$ for fast ($> 900$ km s$^{-1}$) CMEs, and as large as 12\,--\,50 R$_{\odot}$ for slower ones 
(47\,--\,890 km s$^{-1}$). 

\item
The distance $h_{peak}$ at which the Lorentz force peaks is between 1.65 and 2.45 R$_{\odot}$ for all CMEs.

\item
At $\widetilde{h}_{0}$, the Lorentz force has typically fallen by 20\,--\,60 \% (relative to its peak value) for fast CMEs. 
For slower CMEs, the decrease ranges between 70\,--\,98\%.

\item 
Well beyond $\widetilde{h}_{0}$, the drag force exceeds the Lorentz driving force by a significant amount for fast 
CMEs (50\%\,--\,90\%). However, for some slow CMEs the dominance of the drag force is not as pronounced, 
suggesting that part of the CME's magnetic flux may be dissipated in aiding its expansion or heating.
\end{itemize}

In calculating the Lorentz force, the initial equilibrium position for the CME flux rope is is taken 
to be $h_{eq}$ = 1.05 R$_{\odot}$ for all events. The overlying field is taken to decrease as $B_{ex} \propto R^{-n}$. 
The quantity $n$ needs to be greater than a critical value $n_{cr}$ ($n_{cr}=3/2-1/(4c^{'}_{eq})$) for the torus instability 
to be operative, ensuring CME eruption. For each CME, we choose 
a value of $n$ that is $> n_{cr}$. We demand that the Lorentz force equals the aerodynamic drag force at $\widetilde{h}_{0}$. 
The value for $n$ is chosen such that the Lorentz force remains lower than the aerodynamic drag force beyond $\widetilde{h}_{0}$.
For the CMEs in our sample, the critical decay index $n_{cr}$ ranges from 1.29 to 1.39.

For a fixed value of $n$, we note that an increase in $h_{eq}$ by 14 $\%$ 
increases the peak force position value by $\sim 15 \%$. It decreases the $Fall\,\%$ of the Lorentz force 
at $\widetilde{h}_{0}$ (relative to its peak value) by 5 $\%$.
For a fixed value of $h_{eq} (-1.05$ R$_{\odot})$, an increase in n by $31 \%$ decreases the peak position by 17 $\%$. 
It also increases the $Fall\,\%$ of the Lorentz force at $\widetilde{h}_{0}$ (relative to its peak value) by 19.5 $\%$.

Although we have considered only Lorentz and solar wind aerodynamic drag in order to explain CME dynamics, 
we note that there can be other important contributors to the overall energetics. For instance, the work involved 
in CME expansion and the energy expended in possibly heating the CME plasma \textit{e.g.} \citet{Kum96,Wan09,Ems12}
are not considered here. These could well be important, in addition to the energy dissipated due to aerodynamic drag. 
An understanding of these quantities can be achieved via observations of CME expansion as well as measurements of 
thermodynamic quantities inside the CME as it progresses through the heliosphere. 
The latter can possibly be done with the upcoming \textit{Solar Probe Plus} and \textit{Solar Orbiter} missions or via an off-limb spectroscopy mission.


{\bf Acknowledgements}:
NS acknowledges support from a PhD studentship at IISER Pune, from NAMASTE India-EU scholarship and from the Infosys Foundation 
Travel Award. NS is thankful to A. Pluta and N. Mrotzek for their help and useful discussions about data selection and fitting procedures.
PS acknowledges support from the Indian Space Research Organization via a RESPOND grant. 
AV is supported by NNX16AH70G. VB acknowledges support of the CGAUSS (Coronagraphic German and US \textit{Solar Probe Plus} Survey) project 
for WISPR by the German Space Agency DLR under grant 50 OL 1601. 
The SECCHI data are produced by an international consortium of the NRL, LMSAL and NASA/GSFC (USA), RAL and 
Univ. Bham (UK), MPS (Germany), CSL (Belgium), IOTA and IAS (France).


\begin{table}

\caption{Details of all the CMEs in the sample. Near-Earth and observational GCS parameters.
The first column is the serial number of each event with which it is referenced in the article.
For each event the observation date and time when it is first fitted in C2 FOV is shown in the second and third columns.
$h_{0}$ is the observed GCS height at the first observation and $v_{0}$ is the derived velocity at 
$h_{0}$. $n_{wind}$ and $v_{wind}$ are the observed proton number density and solar wind speed at 1 AU respectively.
GCS parameters at $h_{0}$ are given by the Carrington longitude, $\phi$, heliographic latitude, $\theta$, 
tilt, $\gamma$, aspect ratio, $\kappa$, and half angle, $\alpha$. All the fast CMEs are indicated by a 
superscript (f) in their serial number. The events from Paper 1 are indicated by a superscript($\ast$) 
by their corresponding serial number.}
\label{tbl1}
 \begin{tabular}{lcccccccccc} 
  \hline
       &     &  &    &  &  & \multicolumn{5}{c}{GCS Parameters at $h_{0}$} \\
 No.& Date\hspace{1.0cm}Time & $h_{0}$ &  $v_{0}$& $n_{wind}$& $v_{wind}$& $\phi$&$\theta$& $\gamma$ & $\kappa$
 &$\alpha$  \\
    &\hspace{1.7cm}[UT]&[R$_{\odot}$]&[km s$^{-1}$]&[cm$^{-3}$]&[km s$^{-1}$] &[$^{\circ}$]&[$^{\circ}$]&
    [$^{\circ}$]& &[$^{\circ}$]\\
  \hline
  1$^{*}$ & 2010 Mar. 19 11:39 & 3.5 & 162 & 3.6 & 380 & 119 & -10 & -35 & 0.28 & 10 \\
  2$^{* f}$ & 2010 Apr. 03 10:24 & 5.5 & 916 & 7.1 & 470 & 267 & -25 &  33 & 0.34 & 25 \\
  3$^{*}$   & 2010 Apr. 08 03:24 & 2.9 & 468 & 3.60 & 440 & 180 &  17 & -18 & 0.20 & 22 \\
  4$^{*}$   & 2010 Jun. 16 15:24 & 5.7 & 193 &3.50 & 500  & 336 & 0.5 & -15 & 0.23 & 9.5 \\
  5$^{*}$   & 2010 Sep. 11 02:24 & 4.0 & 444 & 4.00 & 320 & 260 & 23  & -49 & 0.41 & 18 \\
  6$^{*}$   & 2010 Oct. 26 07:39 & 5.3 & 215 & 3.80 & 350 & 74  & -31 & -55 & 0.25 & 22\\
  7         & 2010 Dec. 23 05:54 & 3.7 & 147 & 6.10 & 321 & 29  & -28 & -15 & 0.40 & 18 \\
  8         & 2011 Jan. 24 03:54 & 4.4 & 276 & 9.00 & 320 & 336 & -15 & -15 & 0.30 & 22 \\
  9$^{*}$   & 2011 Feb. 15 02:24 & 4.4 & 832 & 2.50 & 440 & 30  & -6  &  30 & 0.47 & 27 \\
  10        & 2011 Mar. 03 05:54 & 4.9 & 349 & 2.25 & 550 & 175 & -22 &  8  & 0.35 & 21 \\
  11$^{*}$  & 2011 Mar. 25 07:00 & 4.8 & 47  & 3.00 & 360 & 207 &   1 &  9  & 0.21 & 37 \\
  12        & 2011 Apr. 08 23:39 & 4.7 & 300 & 5.00 & 375 &  41 &   6 & -6  & 0.30 & 35 \\
  13        & 2011 Jun. 14 07:24 & 3.6 & 562 & 3.70 & 455 & 202 & 1   & 36  & 0.26 & 57\\
  14$^{f}$  & 2011 Jun. 21 03:54 & 8.4 & 1168 & 8.00& 470 & 129 & 5   & -8  & 0.45 & 14\\
  15$^{f}$  & 2011 Jul. 09 00:54 & 4.1 & 903 & 7.50 & 445 & 264 & 17  & 15  & 0.35 & 18 \\
  16$^{f}$  & 2011 Aug. 04 04:24 & 7.3 & 1638& 2.00 & 355 & 324 & 19  & 65  & 0.69 & 29 \\
  17        & 2011 Sep. 13 23:39 & 3.8 & 493 & 2.13 & 468 & 134 & 19  & -38 & 0.43 & 41\\
  18$^{f}$  & 2011 Oct. 22 10:54 & 4.0 & 1276& 8.00 & 300 & 54  & 44  & 16  & 0.60 & 45\\
  19        & 2011 Oct. 26 12:39 & 7.8 & 889 & 3.00 & 260 & 302 & 7   & -1  & 0.46 & 9\\
  20        & 2011 Oct. 27 12:39 & 5.3 & 882 & 8.42 & 411 & 223 & 29  & 16  & 0.36 & 16 \\
  21$^{f}$  & 2012 Jan. 19 15:24 & 4.6 & 1823& 7.00 & 310 & 212 & 44  & 90  & 0.47 & 58\\
  22$^{f}$  & 2012 Jan. 23 03:24 & 4.0 & 1910& 6.00 & 416 & 206 & 28  & 58  & 0.48 & 41\\
  23$^{f}$  & 2012 Jan. 27 17:54 & 3.5 & 2397& 4.00 & 420 & 193 & 30  & 69  & 0.38 & 41\\
  24$^{f}$  & 2012 Mar. 13 17:39 & 3.9 & 1837& 1.00 & 533 & 302 & 21  & -40 & 0.74 & 73 \\
  25        & 2012 Apr. 19 15:39 & 4.1 & 648 & 10.00& 325 & 82  & -28 & 0.0 & 0.27 & 30 \\
  26$^{f}$  & 2012 Jun. 14 14:24 & 6.2 & 1152& 3.23 & 324 & 92  & -22 & -87 & 0.38 & 20\\
  27$^{f}$  & 2012 Jul. 12 16:54 & 4.4 & 1248& 3.20 & 355 & 88  & -10 & 78  & 0.45 & 35 \\
  28$^{f}$  & 2012 Sep. 28 00:24 & 6.7 & 1305& 7.00 & 320 & 165 & 17  & 86  & 0.42 & 42\\
  29        & 2012 Oct. 05 03:39 & 4.4 & 461 & 6.00 & 320 & 56  & -24 & 37  & 0.30 & 31\\
  30        & 2012 Oct. 27 17:24 & 7.3 & 380 & 5.00 & 280 & 118 & 8   & -36 & 0.20 & 40 \\
  31        & 2012 Nov. 09 14:54 & 3.8 & 602 & 13.00& 290 & 285 & -18 &  7  & 0.48 & 35 \\
  32        & 2012 Nov. 23 14:39 & 6.3 & 492 & 7.00 & 370 & 91  & -21 & -66 & 0.52 & 10\\
  33$^{f}$  & 2013 Mar. 15 06:54 & 4.7 & 1504& 4.50 & 470 & 76  & -7  & -86 & 0.31 & 40\\
  34$^{f}$  & 2013 Apr. 11 07:39 & 5.9 & 1115& 3.30 & 445 & 77  & -1  & 90  & 0.14 & 47\\
  35$^{f}$  & 2013 Jun. 28 02:24 & 6.6 & 1637& 10.00& 420 & 177 & -35 & -20 & 0.41 &  5 \\
  36$^{f}$  & 2013 Sep. 29 22:24 & 4.9 & 1217& 11.00& 260 & 360 & 21  & 90  & 0.38 & 47\\
  37$^{f}$  & 2013 Nov. 07 00:24 & 5.9 & 975 & 5.50 & 381 & 304 & -30 & -75 & 0.34 & 12\\
  38$^{f}$  & 2013 Dec. 07 08:24 & 6.8 & 1039& 15.00& 367 & 221 & 32  & 51  & 0.36 & 47 \\

\hline
\end{tabular}
\end{table}

\begin{table}
\caption{Parameters for the solar wind drag- and Lorentz-force analysis are shown here. The first column indicates the serial number of the
CME from Table \ref{tbl1}. $\widetilde{h}_{0}$ is the height at which drag force takes over the CME dynamics and $\widetilde{v}_{0}$ is the corresponding speed at this height. Lorentz force parameters include the decay index, $n$, height at which the Lorentz 
force peaks ($h_{peak}$), equilibrium current, $I_{eq}$, at $h_{eq}$, equilibrium field, $B_{ext}$, also evaluated at $R=h_{eq}$ and $Fall\, \%$ which
gives the amount by which Lorentz force decreases from its maximum value (at $h_{peak}$) to its value at $\widetilde{h}_{0}$. In the last column, $F_{\rm diff}$ denotes the quantity $\frac{F_{\rm drag}-F_{\rm Lorentz}}{F_{\rm drag}}\times 100 \%$ at 
40 R$_{\odot}$ for all events (except CME 11, for which it is evaluated at 50 R$_{\odot}$).}
\label{tbl2}
 \begin{tabular}{lcccccccc} 
  \hline
   & \multicolumn{2}{c}{Drag Parameters}      & \multicolumn{5}{c}{Lorentz Force Parameters} \\
CME    & $\widetilde{h}_{0}$ & $\widetilde{v}_{0}$ &  $n$  & $h_{peak}$ &$I_{eq}$ & $B_{ext}(h_{eq})$&  $Fall \,\%$ & $F_{\rm diff}$\\
No.  &[R$_{\odot}$] &[km s$^{-1}$]& & [R$_{\odot}$]&[10$^{10}$ A] & [10$^{-1}$ G] &  [$\%$] & [$\%$]\\
  \hline
  1$^{*}$   & 21.9 & 383 &  2.5 & 1.75 &  0.41 &0.13& 96 & 18.6\\
  2$^{* f}$ & 5.5  & 916 &  1.6 & 2.35 &  3.13 &0.94& 30 & 43.2\\
  3$^{*}$   & 19.7 & 506 &  1.9 & 2.05 &  0.55 &0.19& 86 & 16.3\\
  4$^{*}$   & 15.2 & 437 &  2.5 & 1.75 &  0.31 &0.11& 93 & -41.3\\
  5$^{*}$   & 27.7 & 490 &  1.6 & 2.35 &  1.77 &0.33& 79 & 6.6\\
  6$^{*}$   & 20.1 & 445 &  1.7 & 2.25 &  0.66 &0.22& 73 & 5.9\\
  7         & 27.1 & 583 &  1.6 & 2.35 &  2.30 &0.65& 81 & 4.2\\
  8         & 20.8 & 454 &  1.6 & 2.35 &  1.21 &0.38& 77 & 19.7\\
  9$^{*}$   & 39.7 & 530 &  2.1 & 1.95 &  1.10 &0.29& 97 & 0.2\\
  10        & 18.2 & 511 &  2.5 & 1.75 &  0.50 &0.15& 95 & -33.3\\
  11$^{*}$  & 46.5 & 456 &  1.9 & 2.05 &  0.71 &0.25& 94 & 0.9\\
  12        & 12.1 & 373 &  2.5 & 1.75 &  0.47 &0.15& 91 & 24.3\\
  13	    & 24.4 & 767 &  1.6 & 2.35 &  1.72 &0.56& 80 & 30.7\\
  14$^{f}$  & 8.4  & 1168&  1.6 & 2.35 &  6.26 &1.71& 48 & 21.6\\
  15$^{f}$  & 4.1  & 903 &  1.9 & 2.05 &  2.66 &0.79& 29 & 52.4\\
  16$^{f}$  & 7.3  & 1638&  1.6 & 2.45 &  5.90 &1.39& 41 & 61.3\\
  17        & 38.8 & 636 &  1.7 & 2.25 &  1.06 &0.29& 91 & 0.3\\
  18$^{f}$  & 4.0  & 1276&  2.1 & 1.95 &  8.40 &2.09& 35 & 80.0\\
  19        & 30.5 & 313 &  2.1 & 1.95 &  0.47 &0.13& 96 & 2.1\\
  20        & 39.4 & 491 &  2.2 & 1.95 &  1.67 &0.49& 98 & 0.3\\
  21$^{f}$  & 4.6  & 1823&  3.0 & 1.65 &  11.60&3.11& 66 & 80.9\\
  22$^{f}$  & 4.0  & 1910&  3.0 & 1.65 &  10.30&2.74& 58 & 93.7\\
  23$^{f}$  & 3.5  & 2397&  3.0 & 1.65 &  8.51 &2.47& 49 & 94.6\\
  24$^{f}$  & 3.9  & 1837&  1.9 & 2.05 &  3.92 &0.91& 25 & 83.2\\
  25        & 23.1 & 684 &  1.6 & 2.35 &  3.68 &1.19& 71 & 3.3\\
  26$^{f}$  & 6.2  & 1152&  1.6 & 2.35 &  2.89 &0.84& 35 & 70.5\\
  27$^{f}$  & 4.4  & 1248&  1.6 & 2.35 &  4.07 &1.11& 18 & 64.9\\
  28$^{f}$  & 6.7  & 1305&  1.6 & 2.35 &  8.53 &2.37& 39 & 59.2\\
  29        & 31.1 & 790 &  1.6 & 2.35 &  4.05 &1.28& 79 & 7.8\\
  30        & 36.9 & 570 &  1.6 & 2.35 &  1.56 &0.56& 84 & 29.2\\
  31        & 26.5 & 597 &  2.9 & 1.75 &  11.07&2.96& 98 & 4.4\\
  32        & 27.7 & 668 &  1.7 & 2.25 &  3.41 &0.89& 86 & 10.4\\
  33$^{f}$  & 4.7  & 1504&  1.8 & 2.15 &  4.29 &1.32& 32 & 56.0\\
  34$^{f}$  & 5.9  & 1115&  1.6 & 2.35 &  1.29 &0.52& 34 & 83.5\\
  35$^{f}$  & 6.6  & 1637&  2.5 & 1.85 &  9.55 &2.69& 26 & 60.3\\
  36$^{f}$  & 4.9  & 1217&  2.1 & 1.95 &  7.06 &2.04& 48 & 80.7\\
  37$^{f}$  & 5.9  & 975 &  1.7 & 2.25 &  2.50 &0.75& 60 & 68.3\\
  38$^{f}$  & 6.8  & 1039&  1.9 & 2.05 &  6.91 &2.04& 57 &  8.9\\
%

  \hline
\end{tabular}
\end{table}
%
%
\newpage


   \clearpage
\bibliographystyle{spr-mp-sola}
\bibliography{references}  

\begin{thebibliography}{49}
\ifx\bisbn     \undefined \def\bisbn  #1{ISBN #1}\fi
\ifx\binits    \undefined \def\binits#1{#1}\fi
\ifx\bauthor   \undefined \def\bauthor#1{#1}\fi
\ifx\batitle   \undefined \def\batitle#1{#1}\fi
\ifx\bjtitle   \undefined \def\bjtitle#1{\textit{#1}}\fi
\ifx\bvolume   \undefined \def\bvolume#1{\textbf{#1}}\fi
\ifx\byear     \undefined \def\byear#1{#1}\fi
\ifx\bissue    \undefined \def\bissue#1{#1}\fi
\ifx\bfpage    \undefined \def\bfpage#1{#1}\fi
\ifx\blpage    \undefined \def\blpage #1{#1}\fi
\ifx\burl      \undefined \def\burl#1{\textsf{#1}}\fi
\ifx\href      \undefined \def\href#1#2{\textsf{#2}}\fi
\ifx\betal     \undefined \def\betal{\textit{et al.}}\fi
\ifx\bctitle   \undefined \def\bctitle#1{#1}\fi
\ifx\beditor   \undefined \def\beditor#1{#1}\fi
\ifx\bbtitle   \undefined \def\bbtitle#1{\textit{#1}}\fi
\ifx\bedition  \undefined \def\bedition#1{#1}\fi
\ifx\bseriesno \undefined \def\bseriesno#1{\textbf{#1}}\fi
\ifx\blocation \undefined \def\blocation#1{#1}\fi
\ifx\bsertitle \undefined \def\bsertitle#1{\textit{#1}}\fi
\ifx\bsnm      \undefined \def\bsnm#1{#1}\fi
\ifx\bsuffix   \undefined \def\bsuffix#1{#1}\fi
\ifx\bparticle \undefined \def\bparticle#1{#1}\fi
\ifx\barticle  \undefined \def\barticle#1{}\fi
\ifx\binstitute  \undefined \def\binstitute#1{#1}\fi
\ifx\bpublisher  \undefined \def\bpublisher#1{#1}\fi
\ifx\doiurl    \undefined
  \def\doiurl#1{\href{http://dx.doi.org/#1}{\textsf{DOI}}}\fi
\ifx\arxivurl  \undefined
  \def\arxivurl#1{\href{http://arxiv.org/abs/#1}{\textsf{arXiv}}}\fi
\ifx\adsurl    \undefined
  \def\adsurl#1{\href{http://adsabs.harvard.edu/abs/#1}{\textsf{ADS}}}\fi
\ifx\botherref \undefined \def\botherref#1{}\fi
\ifx\url       \undefined \def\url#1{\textsf{#1}}\fi
\ifx\bchapter  \undefined \def\bchapter#1{}\fi
\ifx\bbook     \undefined \def\bbook#1{}\fi
\ifx\bcomment  \undefined \def\bcomment#1{#1}\fi
\ifx\oauthor   \undefined \def\oauthor#1{#1}\fi
\ifx\citeauthoryear \undefined\def \citeauthoryear#1{#1}\fi
\ifx\endbibitem\undefined \def\endbibitem{}\fi
\ifx\bconflocation  \undefined \def\bconflocation#1{#1} \fi

\bibitem[\protect\citeauthoryear{{Bein} \textit{et~al.}}{2011}]{Bei11}
\begin{barticle}
\bauthor{\bsnm{{Bein}}, \binits{B.M.}},
\bauthor{\bsnm{{Berkebile-Stoiser}}, \binits{S.}},
\bauthor{\bsnm{{Veronig}}, \binits{A.M.}},
\bauthor{\bsnm{{Temmer}}, \binits{M.}},
\bauthor{\bsnm{{Muhr}}, \binits{N.}},
\bauthor{\bsnm{{Kienreich}}, \binits{I.}}, \betal:
\byear{2011},
\batitle{{Impulsive Acceleration of Coronal Mass Ejections. I. Statistics and
  Coronal Mass Ejection Source Region Characteristics}}.
\bjtitle{\apj}
\bvolume{738},
\bfpage{191}.
\doiurl{10.1088/0004-637X/738/2/191}.
\adsurl{http://esoads.eso.org/abs/2011ApJ...738..191B}.
\end{barticle}
\endbibitem

\bibitem[\protect\citeauthoryear{{Bosman} \textit{et~al.}}{2012}]{Bos12}
\begin{barticle}
\bauthor{\bsnm{{Bosman}}, \binits{E.}},
\bauthor{\bsnm{{Bothmer}}, \binits{V.}},
\bauthor{\bsnm{{Nistic{\`o}}}, \binits{G.}},
\bauthor{\bsnm{{Vourlidas}}, \binits{A.}},
\bauthor{\bsnm{{Howard}}, \binits{R.A.}},
\bauthor{\bsnm{{Davies}}, \binits{J.A.}}:
\byear{2012},
\batitle{{Three-Dimensional Properties of Coronal Mass Ejections from
  STEREO/SECCHI Observations}}.
\bjtitle{\solphys}
\bvolume{281},
\bfpage{167}.
\doiurl{10.1007/s11207-012-0123-5}.
\adsurl{http://esoads.eso.org/abs/2012SoPh..281..167B}.
\end{barticle}
\endbibitem

\bibitem[\protect\citeauthoryear{{Bothmer} and {Daglis}}{2007}]{Bot07}
\begin{bbook}
\bauthor{\bsnm{{Bothmer}}, \binits{V.}},
\bauthor{\bsnm{{Daglis}}, \binits{I.A.}}:
\byear{2007},
\bbtitle{{Space Weather -- Physics and Effects}}.
\doiurl{10.1007/978-3-540-34578-7}.
\adsurl{http://esoads.eso.org/abs/2007swpe.book.....B}.
\end{bbook}
\endbibitem

\bibitem[\protect\citeauthoryear{{Brueckner} \textit{et~al.}}{1995}]{Bru95}
\begin{barticle}
\bauthor{\bsnm{{Brueckner}}, \binits{G.E.}},
\bauthor{\bsnm{{Howard}}, \binits{R.A.}},
\bauthor{\bsnm{{Koomen}}, \binits{M.J.}},
\bauthor{\bsnm{{Korendyke}}, \binits{C.M.}},
\bauthor{\bsnm{{Michels}}, \binits{D.J.}},
\bauthor{\bsnm{{Moses}}, \binits{J.D.}}, \betal:
\byear{1995},
\batitle{{The Large Angle Spectroscopic Coronagraph (LASCO)}}.
\bjtitle{\solphys}
\bvolume{162},
\bfpage{357}.
\doiurl{10.1007/BF00733434}.
\adsurl{http://esoads.eso.org/abs/1995SoPh..162..357B}.
\end{barticle}
\endbibitem

\bibitem[\protect\citeauthoryear{{Cargill}}{2004}]{Car04}
\begin{barticle}
\bauthor{\bsnm{{Cargill}}, \binits{P.J.}}:
\byear{2004},
\batitle{{On the Aerodynamic Drag Force Acting on Interplanetary Coronal Mass
  Ejections}}.
\bjtitle{\solphys}
\bvolume{221},
\bfpage{135}.
\doiurl{10.1023/B:SOLA.0000033366.10725.a2}.
\adsurl{http://esoads.eso.org/abs/2004SoPh..221..135C}.
\end{barticle}
\endbibitem

\bibitem[\protect\citeauthoryear{{Carley}, {McAteer}, and
  {Gallagher}}{2012}]{Car12}
\begin{barticle}
\bauthor{\bsnm{{Carley}}, \binits{E.P.}},
\bauthor{\bsnm{{McAteer}}, \binits{R.T.J.}},
\bauthor{\bsnm{{Gallagher}}, \binits{P.T.}}:
\byear{2012},
\batitle{{Coronal Mass Ejection Mass, Energy, and Force Estimates Using
  STEREO}}.
\bjtitle{\apj}
\bvolume{752},
\bfpage{36}.
\doiurl{10.1088/0004-637X/752/1/36}.
\adsurl{http://esoads.eso.org/abs/2012ApJ...752...36C}.
\end{barticle}
\endbibitem

\bibitem[\protect\citeauthoryear{{Chen}}{1989}]{Che89}
\begin{barticle}
\bauthor{\bsnm{{Chen}}, \binits{J.}}:
\byear{1989},
\batitle{{Effects of toroidal forces in current loops embedded in a background
  plasma}}.
\bjtitle{\apj}
\bvolume{338},
\bfpage{453}.
\doiurl{10.1086/167211}.
\adsurl{http://esoads.eso.org/abs/1989ApJ...338..453C}.
\end{barticle}
\endbibitem

\bibitem[\protect\citeauthoryear{{Chen}}{1996}]{Che96}
\begin{barticle}
\bauthor{\bsnm{{Chen}}, \binits{J.}}:
\byear{1996},
\batitle{{Theory of prominence eruption and propagation: Interplanetary
  consequences}}.
\bjtitle{\jgr}
\bvolume{101},
\bfpage{27499}.
\doiurl{10.1029/96JA02644}.
\adsurl{http://esoads.eso.org/abs/1996JGR...10127499C}.
\end{barticle}
\endbibitem

\bibitem[\protect\citeauthoryear{{Chen} and {Kunkel}}{2010}]{Che10}
\begin{barticle}
\bauthor{\bsnm{{Chen}}, \binits{J.}},
\bauthor{\bsnm{{Kunkel}}, \binits{V.}}:
\byear{2010},
\batitle{{Temporal and Physical Connection Between Coronal Mass Ejections and
  Flares}}.
\bjtitle{\apj}
\bvolume{717},
\bfpage{1105}.
\doiurl{10.1088/0004-637X/717/2/1105}.
\adsurl{http://esoads.eso.org/abs/2010ApJ...717.1105C}.
\end{barticle}
\endbibitem

\bibitem[\protect\citeauthoryear{{Chen} \textit{et~al.}}{1997}]{Che97}
\begin{barticle}
\bauthor{\bsnm{{Chen}}, \binits{J.}},
\bauthor{\bsnm{{Howard}}, \binits{R.A.}},
\bauthor{\bsnm{{Brueckner}}, \binits{G.E.}},
\bauthor{\bsnm{{Santoro}}, \binits{R.}},
\bauthor{\bsnm{{Krall}}, \binits{J.}},
\bauthor{\bsnm{{Paswaters}}, \binits{S.E.}},
\bauthor{\bsnm{{et. al.}}}:
\byear{1997},
\batitle{{Evidence of an Erupting Magnetic Flux Rope: LASCO Coronal Mass
  Ejection of 1997 April 13}}.
\bjtitle{\apjl}
\bvolume{490},
\bfpage{L191}.
\doiurl{10.1086/311029}.
\adsurl{http://esoads.eso.org/abs/1997ApJ...490L.191C}.
\end{barticle}
\endbibitem

\bibitem[\protect\citeauthoryear{{Emslie} \textit{et~al.}}{2012}]{Ems12}
\begin{barticle}
\bauthor{\bsnm{{Emslie}}, \binits{A.G.}},
\bauthor{\bsnm{{Dennis}}, \binits{B.R.}},
\bauthor{\bsnm{{Shih}}, \binits{A.Y.}},
\bauthor{\bsnm{{Chamberlin}}, \binits{P.C.}},
\bauthor{\bsnm{{Mewaldt}}, \binits{R.A.}},
\bauthor{\bsnm{{Moore}}, \binits{C.S.}}, \betal:
\byear{2012},
\batitle{{Global Energetics of Thirty-eight Large Solar Eruptive Events}}.
\bjtitle{\apj}
\bvolume{759},
\bfpage{71}.
\doiurl{10.1088/0004-637X/759/1/71}.
\adsurl{2012ApJ...759...71E}.
\end{barticle}
\endbibitem

\bibitem[\protect\citeauthoryear{{Forbes}}{2000}]{For00}
\begin{barticle}
\bauthor{\bsnm{{Forbes}}, \binits{T.G.}}:
\byear{2000},
\batitle{{A review on the genesis of coronal mass ejections}}.
\bjtitle{\jgr}
\bvolume{105},
\bfpage{23153}.
\doiurl{10.1029/2000JA000005}.
\adsurl{http://esoads.eso.org/abs/2000JGR...10523153F}.
\end{barticle}
\endbibitem

\bibitem[\protect\citeauthoryear{{Forbes} and {Isenberg}}{1991}]{For91}
\begin{barticle}
\bauthor{\bsnm{{Forbes}}, \binits{T.G.}},
\bauthor{\bsnm{{Isenberg}}, \binits{P.A.}}:
\byear{1991},
\batitle{{A catastrophe mechanism for coronal mass ejections}}.
\bjtitle{\apj}
\bvolume{373},
\bfpage{294}.
\doiurl{10.1086/170051}.
\adsurl{http://esoads.eso.org/abs/1991ApJ...373..294F}.
\end{barticle}
\endbibitem

\bibitem[\protect\citeauthoryear{{Fry} \textit{et~al.}}{2003}]{Fry03}
\begin{barticle}
\bauthor{\bsnm{{Fry}}, \binits{C.D.}},
\bauthor{\bsnm{{Dryer}}, \binits{M.}},
\bauthor{\bsnm{{Smith}}, \binits{Z.}},
\bauthor{\bsnm{{Sun}}, \binits{W.}},
\bauthor{\bsnm{{Deehr}}, \binits{C.S.}},
\bauthor{\bsnm{{Akasofu}}, \binits{S.-I.}}:
\byear{2003},
\batitle{{Forecasting solar wind structures and shock arrival times using an
  ensemble of models}}.
\bjtitle{\jgr}
\bvolume{108},
\bfpage{1070}.
\doiurl{10.1029/2002JA009474}.
\adsurl{http://esoads.eso.org/abs/2003JGRA..108.1070F}.
\end{barticle}
\endbibitem

\bibitem[\protect\citeauthoryear{{Gopalswamy}}{2013}]{Gop13}
\begin{bchapter}
\bauthor{\bsnm{{Gopalswamy}}, \binits{N.}}:
\byear{2013},
\bctitle{{STEREO and SOHO contributions to coronal mass ejection studies: Some
  recent results}}.
In: \bbtitle{Astron. Soc. India C.S.}
\bseriesno{10}.
\adsurl{http://esoads.eso.org/abs/2013ASInC..10...11G}.
\end{bchapter}
\endbibitem

\bibitem[\protect\citeauthoryear{{Gosling} \textit{et~al.}}{1991}]{Gos91}
\begin{barticle}
\bauthor{\bsnm{{Gosling}}, \binits{J.T.}},
\bauthor{\bsnm{{McComas}}, \binits{D.J.}},
\bauthor{\bsnm{{Phillips}}, \binits{J.L.}},
\bauthor{\bsnm{{Bame}}, \binits{S.J.}}:
\byear{1991},
\batitle{{Geomagnetic activity associated with earth passage of interplanetary
  shock disturbances and coronal mass ejections}}.
\bjtitle{\jgr}
\bvolume{96},
\bfpage{7831}.
\doiurl{10.1029/91JA00316}.
\adsurl{1991JGR....96.7831G}.
\end{barticle}
\endbibitem

\bibitem[\protect\citeauthoryear{{Howard} \textit{et~al.}}{2008}]{How08}
\begin{barticle}
\bauthor{\bsnm{{Howard}}, \binits{R.A.}},
\bauthor{\bsnm{{Moses}}, \binits{J.D.}},
\bauthor{\bsnm{{Vourlidas}}, \binits{A.}},
\bauthor{\bsnm{{Newmark}}, \binits{J.S.}},
\bauthor{\bsnm{{Socker}}, \binits{D.G.}},
\bauthor{\bsnm{{Plunkett}}, \binits{S.P.}}, \betal:
\byear{2008},
\batitle{{Sun Earth Connection Coronal and Heliospheric Investigation
  (SECCHI)}}.
\bjtitle{\ssr}
\bvolume{136},
\bfpage{67}.
\doiurl{10.1007/s11214-008-9341-4}.
\adsurl{http://esoads.eso.org/abs/2008SSRv..136...67H}.
\end{barticle}
\endbibitem

\bibitem[\protect\citeauthoryear{{Isenberg} and {Forbes}}{2007}]{Ise07}
\begin{barticle}
\bauthor{\bsnm{{Isenberg}}, \binits{P.A.}},
\bauthor{\bsnm{{Forbes}}, \binits{T.G.}}:
\byear{2007},
\batitle{{A Three-dimensional Line-tied Magnetic Field Model for Solar
  Eruptions}}.
\bjtitle{\apj}
\bvolume{670},
\bfpage{1453}.
\doiurl{10.1086/522025}.
\adsurl{http://esoads.eso.org/abs/2007ApJ...670.1453I}.
\end{barticle}
\endbibitem

\bibitem[\protect\citeauthoryear{{Jin} \textit{et~al.}}{2017}]{Jin17}
\begin{barticle}
\bauthor{\bsnm{{Jin}}, \binits{M.}},
\bauthor{\bsnm{{Manchester}}, \binits{W.B.}},
\bauthor{\bsnm{{van der Holst}}, \binits{B.}},
\bauthor{\bsnm{{Sokolov}}, \binits{I.}},
\bauthor{\bsnm{{T{\'o}th}}, \binits{G.}},
\bauthor{\bsnm{{Mullinix}}, \binits{R.E.}},
\bauthor{\bsnm{{Taktakishvili}}, \binits{A.}},
\bauthor{\bsnm{{Chulaki}}, \binits{A.}},
\bauthor{\bsnm{{Gombosi}}, \binits{T.I.}}:
\byear{2017},
\batitle{{Data-constrained Coronal Mass Ejections in a Global
  Magnetohydrodynamics Model}}.
\bjtitle{\apj}
\bvolume{834},
\bfpage{173}.
\doiurl{10.3847/1538-4357/834/2/173}.
\adsurl{2017ApJ...834..173J}.
\end{barticle}
\endbibitem

\bibitem[\protect\citeauthoryear{{Kaiser} \textit{et~al.}}{2008}]{Kai08}
\begin{barticle}
\bauthor{\bsnm{{Kaiser}}, \binits{M.L.}},
\bauthor{\bsnm{{Kucera}}, \binits{T.A.}},
\bauthor{\bsnm{{Davila}}, \binits{J.M.}},
\bauthor{\bsnm{{St.~Cyr}}, \binits{O.C.}},
\bauthor{\bsnm{{Guhathakurta}}, \binits{M.}},
\bauthor{\bsnm{{Christian}}, \binits{E.}}:
\byear{2008},
\batitle{{The STEREO Mission: An Introduction}}.
\bjtitle{\ssr}
\bvolume{136},
\bfpage{5}.
\doiurl{10.1007/s11214-007-9277-0}.
\adsurl{http://esoads.eso.org/abs/2008SSRv..136....5K}.
\end{barticle}
\endbibitem

\bibitem[\protect\citeauthoryear{{Kliem} and {T{\"o}r{\"o}k}}{2006}]{Kli06}
\begin{barticle}
\bauthor{\bsnm{{Kliem}}, \binits{B.}},
\bauthor{\bsnm{{T{\"o}r{\"o}k}}, \binits{T.}}:
\byear{2006},
\batitle{{Torus Instability}}.
\bjtitle{\prl}
\bvolume{96}(\bissue{25}),
\bfpage{255002}.
\doiurl{10.1103/PhysRevLett.96.255002}.
\adsurl{http://esoads.eso.org/abs/2006PhRvL..96y5002K}.
\end{barticle}
\endbibitem

\bibitem[\protect\citeauthoryear{{Kliem} \textit{et~al.}}{2014}]{Kli14}
\begin{barticle}
\bauthor{\bsnm{{Kliem}}, \binits{B.}},
\bauthor{\bsnm{{Lin}}, \binits{J.}},
\bauthor{\bsnm{{Forbes}}, \binits{T.G.}},
\bauthor{\bsnm{{Priest}}, \binits{E.R.}},
\bauthor{\bsnm{{T{\"o}r{\"o}k}}, \binits{T.}}:
\byear{2014},
\batitle{{Catastrophe versus Instability for the Eruption of a Toroidal Solar
  Magnetic Flux Rope}}.
\bjtitle{\apj}
\bvolume{789},
\bfpage{46}.
\doiurl{10.1088/0004-637X/789/1/46}.
\adsurl{http://esoads.eso.org/abs/2014ApJ...789...46K}.
\end{barticle}
\endbibitem

\bibitem[\protect\citeauthoryear{{Kumar} and {Rust}}{1996}]{Kum96}
\begin{barticle}
\bauthor{\bsnm{{Kumar}}, \binits{A.}},
\bauthor{\bsnm{{Rust}}, \binits{D.M.}}:
\byear{1996},
\batitle{{Interplanetary magnetic clouds, helicity conservation, and
  current-core flux-ropes}}.
\bjtitle{\jgr}
\bvolume{101},
\bfpage{15667}.
\doiurl{10.1029/96JA00544}.
\adsurl{http://esoads.eso.org/abs/1996JGR...10115667K}.
\end{barticle}
\endbibitem

\bibitem[\protect\citeauthoryear{{Lee} \textit{et~al.}}{2013}]{Lee13}
\begin{barticle}
\bauthor{\bsnm{{Lee}}, \binits{C.O.}},
\bauthor{\bsnm{{Arge}}, \binits{C.N.}},
\bauthor{\bsnm{{Odstr{\v c}il}}, \binits{D.}},
\bauthor{\bsnm{{Millward}}, \binits{G.}},
\bauthor{\bsnm{{Pizzo}}, \binits{V.}},
\bauthor{\bsnm{{Quinn}}, \binits{J.M.}}, \betal:
\byear{2013},
\batitle{{Ensemble Modeling of CME Propagation}}.
\bjtitle{\solphys}
\bvolume{285},
\bfpage{349}.
\doiurl{10.1007/s11207-012-9980-1}.
\adsurl{http://esoads.eso.org/abs/2013SoPh..285..349L}.
\end{barticle}
\endbibitem

\bibitem[\protect\citeauthoryear{{Lugaz} \textit{et~al.}}{2007}]{Lug07}
\begin{barticle}
\bauthor{\bsnm{{Lugaz}}, \binits{N.}},
\bauthor{\bsnm{{Manchester}}, \binits{W.B.} \bsuffix{IV}},
\bauthor{\bsnm{{Roussev}}, \binits{I.I.}},
\bauthor{\bsnm{{T{\'o}th}}, \binits{G.}},
\bauthor{\bsnm{{Gombosi}}, \binits{T.I.}}:
\byear{2007},
\batitle{{Numerical Investigation of the Homologous Coronal Mass Ejection
  Events from Active Region 9236}}.
\bjtitle{\apj}
\bvolume{659},
\bfpage{788}.
\doiurl{10.1086/512005}.
\adsurl{http://esoads.eso.org/abs/2007ApJ...659..788L}.
\end{barticle}
\endbibitem

\bibitem[\protect\citeauthoryear{{Mays} \textit{et~al.}}{2015}]{May15}
\begin{barticle}
\bauthor{\bsnm{{Mays}}, \binits{M.L.}},
\bauthor{\bsnm{{Taktakishvili}}, \binits{A.}},
\bauthor{\bsnm{{Pulkkinen}}, \binits{A.}},
\bauthor{\bsnm{{MacNeice}}, \binits{P.J.}},
\bauthor{\bsnm{{Rast{\"a}tter}}, \binits{L.}},
\bauthor{\bsnm{{Odstrcil}}, \binits{D.}},
\bauthor{\bsnm{{Jian}}, \binits{L.K.}},
\bauthor{\bsnm{{Richardson}}, \binits{I.G.}},
\bauthor{\bsnm{{LaSota}}, \binits{J.A.}},
\bauthor{\bsnm{{Zheng}}, \binits{Y.}},
\bauthor{\bsnm{{Kuznetsova}}, \binits{M.M.}}:
\byear{2015},
\batitle{{Ensemble Modeling of CMEs Using the WSA-ENLIL+Cone Model}}.
\bjtitle{\solphys}
\bvolume{290},
\bfpage{1775}.
\doiurl{10.1007/s11207-015-0692-1}.
\adsurl{http://esoads.eso.org/abs/2015SoPh..290.1775M}.
\end{barticle}
\endbibitem

\bibitem[\protect\citeauthoryear{{McKenna-Lawlor}
  \textit{et~al.}}{2006}]{McK06}
\begin{barticle}
\bauthor{\bsnm{{McKenna-Lawlor}}, \binits{S.M.P.}},
\bauthor{\bsnm{{Dryer}}, \binits{M.}},
\bauthor{\bsnm{{Kartalev}}, \binits{M.D.}},
\bauthor{\bsnm{{Smith}}, \binits{Z.}},
\bauthor{\bsnm{{Fry}}, \binits{C.D.}},
\bauthor{\bsnm{{Sun}}, \binits{W.}}, \betal:
\byear{2006},
\batitle{{Near real-time predictions of the arrival at Earth of flare-related
  shocks during Solar Cycle 23}}.
\bjtitle{\jgr}
\bvolume{111},
\bfpage{A11103}.
\doiurl{10.1029/2005JA011162}.
\adsurl{http://esoads.eso.org/abs/2006JGRA..11111103M}.
\end{barticle}
\endbibitem

\bibitem[\protect\citeauthoryear{{Mishra} and {Srivastava}}{2013}]{Mis13}
\begin{barticle}
\bauthor{\bsnm{{Mishra}}, \binits{W.}},
\bauthor{\bsnm{{Srivastava}}, \binits{N.}}:
\byear{2013},
\batitle{{Estimating the Arrival Time of Earth-directed Coronal Mass Ejections
  at in Situ Spacecraft Using COR and HI Observations from STEREO}}.
\bjtitle{\apj}
\bvolume{772},
\bfpage{70}.
\doiurl{10.1088/0004-637X/772/1/70}.
\adsurl{http://esoads.eso.org/abs/2013ApJ...772...70M}.
\end{barticle}
\endbibitem

\bibitem[\protect\citeauthoryear{{Rollett} \textit{et~al.}}{2016}]{Rol16}
\begin{barticle}
\bauthor{\bsnm{{Rollett}}, \binits{T.}},
\bauthor{\bsnm{{M{\"o}stl}}, \binits{C.}},
\bauthor{\bsnm{{Isavnin}}, \binits{A.}},
\bauthor{\bsnm{{Davies}}, \binits{J.A.}},
\bauthor{\bsnm{{Kubicka}}, \binits{M.}},
\bauthor{\bsnm{{Amerstorfer}}, \binits{U.V.}}, \betal:
\byear{2016},
\batitle{{ElEvoHI: A Novel CME Prediction Tool for Heliospheric Imaging
  Combining an Elliptical Front with Drag-based Model Fitting}}.
\bjtitle{\apj}
\bvolume{824},
\bfpage{131}.
\doiurl{10.3847/0004-637X/824/2/131}.
\adsurl{http://esoads.eso.org/abs/2016ApJ...824..131R}.
\end{barticle}
\endbibitem

\bibitem[\protect\citeauthoryear{{Sachdeva} \textit{et~al.}}{2015}]{Sac15}
\begin{barticle}
\bauthor{\bsnm{{Sachdeva}}, \binits{N.}},
\bauthor{\bsnm{{Subramanian}}, \binits{P.}},
\bauthor{\bsnm{{Colaninno}}, \binits{R.}},
\bauthor{\bsnm{{Vourlidas}}, \binits{A.}}:
\byear{2015},
\batitle{{CME Propagation: Where does Aerodynamic Drag 'Take Over'?}}
\bjtitle{\apj}
\bvolume{809},
\bfpage{158}.
\doiurl{10.1088/0004-637X/809/2/158}.
\adsurl{http://esoads.eso.org/abs/2015ApJ...809..158S}.
\end{barticle}
\endbibitem

\bibitem[\protect\citeauthoryear{{Shafranov}}{1966}]{Sha66}
\begin{barticle}
\bauthor{\bsnm{{Shafranov}}, \binits{V.D.}}:
\byear{1966},
\batitle{{Plasma Equilibrium in a Magnetic Field}}.
\bjtitle{Rev. Plasma Phys.}
\bvolume{2},
\bfpage{103}.
\adsurl{http://esoads.eso.org/abs/1966RvPP....2..103S}.
\end{barticle}
\endbibitem

\bibitem[\protect\citeauthoryear{{Subramanian} and {Vourlidas}}{2007}]{Sub07}
\begin{barticle}
\bauthor{\bsnm{{Subramanian}}, \binits{P.}},
\bauthor{\bsnm{{Vourlidas}}, \binits{A.}}:
\byear{2007},
\batitle{{Energetics of solar coronal mass ejections}}.
\bjtitle{\aap}
\bvolume{467},
\bfpage{685}.
\doiurl{10.1051/0004-6361:20066770}.
\adsurl{http://esoads.eso.org/abs/2007A\%26A...467..685S}.
\end{barticle}
\endbibitem

\bibitem[\protect\citeauthoryear{{Subramanian}, {Lara}, and
  {Borgazzi}}{2012}]{Sub12}
\begin{barticle}
\bauthor{\bsnm{{Subramanian}}, \binits{P.}},
\bauthor{\bsnm{{Lara}}, \binits{A.}},
\bauthor{\bsnm{{Borgazzi}}, \binits{A.}}:
\byear{2012},
\batitle{{Can solar wind viscous drag account for coronal mass ejection
  deceleration?}}
\bjtitle{\grl}
\bvolume{39},
\bfpage{L19107}.
\doiurl{10.1029/2012GL053625}.
\adsurl{http://esoads.eso.org/abs/2012GeoRL..3919107S}.
\end{barticle}
\endbibitem

\bibitem[\protect\citeauthoryear{{Taktakishvili} \textit{et~al.}}{2009}]{Tak09}
\begin{barticle}
\bauthor{\bsnm{{Taktakishvili}}, \binits{A.}},
\bauthor{\bsnm{{Kuznetsova}}, \binits{M.}},
\bauthor{\bsnm{{MacNeice}}, \binits{P.}},
\bauthor{\bsnm{{Hesse}}, \binits{M.}},
\bauthor{\bsnm{{Rast{\"a}tter}}, \binits{L.}},
\bauthor{\bsnm{{Pulkkinen}}, \binits{A.}}, \betal:
\byear{2009},
\batitle{{Validation of the coronal mass ejection predictions at the Earth
  orbit estimated by ENLIL heliosphere cone model}}.
\bjtitle{Space Weather}
\bvolume{7},
\bfpage{S03004}.
\doiurl{10.1029/2008SW000448}.
\adsurl{http://esoads.eso.org/abs/2009SpWea...7.3004T}.
\end{barticle}
\endbibitem

\bibitem[\protect\citeauthoryear{{Temmer} and {Nitta}}{2015}]{Tem15}
\begin{barticle}
\bauthor{\bsnm{{Temmer}}, \binits{M.}},
\bauthor{\bsnm{{Nitta}}, \binits{N.V.}}:
\byear{2015},
\batitle{{Interplanetary Propagation Behavior of the Fast Coronal Mass Ejection
  on 23 July 2012}}.
\bjtitle{\solphys}
\bvolume{290},
\bfpage{919}.
\doiurl{10.1007/s11207-014-0642-3}.
\adsurl{http://esoads.eso.org/abs/2015SoPh..290..919T}.
\end{barticle}
\endbibitem

\bibitem[\protect\citeauthoryear{{Temmer} \textit{et~al.}}{2012}]{Tem12}
\begin{barticle}
\bauthor{\bsnm{{Temmer}}, \binits{M.}},
\bauthor{\bsnm{{Vr{\v s}nak}}, \binits{B.}},
\bauthor{\bsnm{{Rollett}}, \binits{T.}},
\bauthor{\bsnm{{Bein}}, \binits{B.}},
\bauthor{\bsnm{{de Koning}}, \binits{C.A.}},
\bauthor{\bsnm{{Liu}}, \binits{Y.}},
\bauthor{\bsnm{{Bosman}}, \binits{E.}},
\bauthor{\bsnm{{Davies}}, \binits{J.A.}},
\bauthor{\bsnm{{M{\"o}stl}}, \binits{C.}},
\bauthor{\bsnm{{{\v Z}ic}}, \binits{T.}},
\bauthor{\bsnm{{Veronig}}, \binits{A.M.}},
\bauthor{\bsnm{{Bothmer}}, \binits{V.}},
\bauthor{\bsnm{{Harrison}}, \binits{R.}},
\bauthor{\bsnm{{Nitta}}, \binits{N.}},
\bauthor{\bsnm{{Bisi}}, \binits{M.}},
\bauthor{\bsnm{{Flor}}, \binits{O.}},
\bauthor{\bsnm{{Eastwood}}, \binits{J.}},
\bauthor{\bsnm{{Odstrcil}}, \binits{D.}},
\bauthor{\bsnm{{Forsyth}}, \binits{R.}}:
\byear{2012},
\batitle{{Characteristics of Kinematics of a Coronal Mass Ejection during the
  2010 August 1 CME-CME Interaction Event}}.
\bjtitle{\apj}
\bvolume{749},
\bfpage{57}.
\doiurl{10.1088/0004-637X/749/1/57}.
\adsurl{http://esoads.eso.org/abs/2012ApJ...749...57T}.
\end{barticle}
\endbibitem

\bibitem[\protect\citeauthoryear{{Thernisien}}{2011}]{The11}
\begin{barticle}
\bauthor{\bsnm{{Thernisien}}, \binits{A.}}:
\byear{2011},
\batitle{{Implementation of the Graduated Cylindrical Shell Model for the
  Three-dimensional Reconstruction of Coronal Mass Ejections}}.
\bjtitle{\apjs}
\bvolume{194},
\bfpage{33}.
\doiurl{10.1088/0067-0049/194/2/33}.
\adsurl{http://esoads.eso.org/abs/2011ApJS..194...33T}.
\end{barticle}
\endbibitem

\bibitem[\protect\citeauthoryear{{Thernisien}, {Vourlidas}, and
  {Howard}}{2009}]{The09}
\begin{barticle}
\bauthor{\bsnm{{Thernisien}}, \binits{A.}},
\bauthor{\bsnm{{Vourlidas}}, \binits{A.}},
\bauthor{\bsnm{{Howard}}, \binits{R.A.}}:
\byear{2009},
\batitle{{Forward Modeling of Coronal Mass Ejections Using STEREO/SECCHI
  Data}}.
\bjtitle{\solphys}
\bvolume{256},
\bfpage{111}.
\doiurl{10.1007/s11207-009-9346-5}.
\adsurl{http://esoads.eso.org/abs/2009SoPh..256..111T}.
\end{barticle}
\endbibitem

\bibitem[\protect\citeauthoryear{{Thernisien}, {Howard}, and
  {Vourlidas}}{2006}]{The06}
\begin{barticle}
\bauthor{\bsnm{{Thernisien}}, \binits{A.F.R.}},
\bauthor{\bsnm{{Howard}}, \binits{R.A.}},
\bauthor{\bsnm{{Vourlidas}}, \binits{A.}}:
\byear{2006},
\batitle{{Modeling of Flux Rope Coronal Mass Ejections}}.
\bjtitle{\apj}
\bvolume{652},
\bfpage{763}.
\doiurl{10.1086/508254}.
\adsurl{http://esoads.eso.org/abs/2006ApJ...652..763T}.
\end{barticle}
\endbibitem

\bibitem[\protect\citeauthoryear{{T{\'o}th} \textit{et~al.}}{2007}]{Tot07}
\begin{barticle}
\bauthor{\bsnm{{T{\'o}th}}, \binits{G.}},
\bauthor{\bsnm{{de Zeeuw}}, \binits{D.L.}},
\bauthor{\bsnm{{Gombosi}}, \binits{T.I.}},
\bauthor{\bsnm{{Manchester}}, \binits{W.B.}},
\bauthor{\bsnm{{Ridley}}, \binits{A.J.}},
\bauthor{\bsnm{{Sokolov}}, \binits{I.V.}}, \betal:
\byear{2007},
\batitle{{Sun-to-thermosphere simulation of the 28-30 October 2003 storm with
  the Space Weather Modeling Framework}}.
\bjtitle{Space Weather}
\bvolume{5},
\bfpage{06003}.
\doiurl{10.1029/2006SW000272}.
\adsurl{http://esoads.eso.org/abs/2007SpWea...5.6003T}.
\end{barticle}
\endbibitem

\bibitem[\protect\citeauthoryear{{Vr{\v s}nak}}{2006}]{Vrs06}
\begin{barticle}
\bauthor{\bsnm{{Vr{\v s}nak}}, \binits{B.}}:
\byear{2006},
\batitle{{Forces governing coronal mass ejections}}.
\bjtitle{\adv}
\bvolume{38},
\bfpage{431}.
\doiurl{10.1016/j.asr.2005.03.090}.
\adsurl{http://esoads.eso.org/abs/2006AdSpR..38..431V}.
\end{barticle}
\endbibitem

\bibitem[\protect\citeauthoryear{{Vr{\v s}nak} \textit{et~al.}}{2010}]{Vrs10}
\begin{barticle}
\bauthor{\bsnm{{Vr{\v s}nak}}, \binits{B.}},
\bauthor{\bsnm{{{\v Z}ic}}, \binits{T.}},
\bauthor{\bsnm{{Falkenberg}}, \binits{T.V.}},
\bauthor{\bsnm{{M{\"o}stl}}, \binits{C.}},
\bauthor{\bsnm{{Vennerstrom}}, \binits{S.}},
\bauthor{\bsnm{{Vrbanec}}, \binits{D.}}:
\byear{2010},
\batitle{{The role of aerodynamic drag in propagation of interplanetary coronal
  mass ejections}}.
\bjtitle{\aap}
\bvolume{512},
\bfpage{A43}.
\doiurl{10.1051/0004-6361/200913482}.
\adsurl{http://esoads.eso.org/abs/2010A\%26A...512A..43V}.
\end{barticle}
\endbibitem

\bibitem[\protect\citeauthoryear{{Vr{\v s}nak} \textit{et~al.}}{2014}]{Vrs14}
\begin{barticle}
\bauthor{\bsnm{{Vr{\v s}nak}}, \binits{B.}},
\bauthor{\bsnm{{Temmer}}, \binits{M.}},
\bauthor{\bsnm{{{\v Z}ic}}, \binits{T.}},
\bauthor{\bsnm{{Taktakishvili}}, \binits{A.}},
\bauthor{\bsnm{{Dumbovi{\'c}}}, \binits{M.}},
\bauthor{\bsnm{{M{\"o}stl}}, \binits{C.}}, \betal:
\byear{2014},
\batitle{{Heliospheric Propagation of Coronal Mass Ejections: Comparison of
  Numerical WSA-ENLIL+Cone Model and Analytical Drag-based Model}}.
\bjtitle{\apjs}
\bvolume{213},
\bfpage{21}.
\doiurl{10.1088/0067-0049/213/2/21}.
\adsurl{http://esoads.eso.org/abs/2014ApJS..213...21V}.
\end{barticle}
\endbibitem

\bibitem[\protect\citeauthoryear{{Wang}, {Zhang}, and {Shen}}{2009}]{Wan09}
\begin{barticle}
\bauthor{\bsnm{{Wang}}, \binits{Y.}},
\bauthor{\bsnm{{Zhang}}, \binits{J.}},
\bauthor{\bsnm{{Shen}}, \binits{C.}}:
\byear{2009},
\batitle{{An analytical model probing the internal state of coronal mass
  ejections based on observations of their expansions and propagations}}.
\bjtitle{\jgr}
\bvolume{114},
\bfpage{A10104}.
\doiurl{10.1029/2009JA014360}.
\adsurl{http://esoads.eso.org/abs/2009JGRA..11410104W}.
\end{barticle}
\endbibitem

\bibitem[\protect\citeauthoryear{{Wood} \textit{et~al.}}{1997}]{Woo99}
\begin{bchapter}
\bauthor{\bsnm{{Wood}}, \binits{B.E.}},
\bauthor{\bsnm{{Karovska}}, \binits{M.}},
\bauthor{\bsnm{{Cook}}, \binits{J.W.}},
\bauthor{\bsnm{{Brueckner}}, \binits{G.E.}},
\bauthor{\bsnm{{Howard}}, \binits{R.A.}}:
\byear{1997},
\bctitle{{LASCO Observations of Variability in the Quiescent Solar Corona}}.
In: \bbtitle{Bull. Amer. Astron. Soc.}
\bseriesno{29},
\bfpage{1321}.
\adsurl{http://esoads.eso.org/abs/1997AAS...191.7303W}.
\end{bchapter}
\endbibitem

\bibitem[\protect\citeauthoryear{{Wu} \textit{et~al.}}{2007}]{Wu07}
\begin{barticle}
\bauthor{\bsnm{{Wu}}, \binits{C.-C.}},
\bauthor{\bsnm{{Fry}}, \binits{C.D.}},
\bauthor{\bsnm{{Wu}}, \binits{S.T.}},
\bauthor{\bsnm{{Dryer}}, \binits{M.}},
\bauthor{\bsnm{{Liou}}, \binits{K.}}:
\byear{2007},
\batitle{{Three-dimensional global simulation of interplanetary coronal mass
  ejection propagation from the Sun to the heliosphere: Solar event of 12 May
  1997}}.
\bjtitle{\jgr}
\bvolume{112},
\bfpage{A09104}.
\doiurl{10.1029/2006JA012211}.
\adsurl{http://esoads.eso.org/abs/2007JGRA..112.9104W}.
\end{barticle}
\endbibitem

\bibitem[\protect\citeauthoryear{{Zhang} and {Dere}}{2006}]{Zha06}
\begin{barticle}
\bauthor{\bsnm{{Zhang}}, \binits{J.}},
\bauthor{\bsnm{{Dere}}, \binits{K.P.}}:
\byear{2006},
\batitle{{A Statistical Study of Main and Residual Accelerations of Coronal
  Mass Ejections}}.
\bjtitle{\apj}
\bvolume{649},
\bfpage{1100}.
\doiurl{10.1086/506903}.
\adsurl{http://esoads.eso.org/abs/2006ApJ...649.1100Z}.
\end{barticle}
\endbibitem

\bibitem[\protect\citeauthoryear{{Zhang} \textit{et~al.}}{2001}]{Zha01}
\begin{barticle}
\bauthor{\bsnm{{Zhang}}, \binits{J.}},
\bauthor{\bsnm{{Dere}}, \binits{K.P.}},
\bauthor{\bsnm{{Howard}}, \binits{R.A.}},
\bauthor{\bsnm{{Kundu}}, \binits{M.R.}},
\bauthor{\bsnm{{White}}, \binits{S.M.}}:
\byear{2001},
\batitle{{On the Temporal Relationship between Coronal Mass Ejections and
  Flares}}.
\bjtitle{\apj}
\bvolume{559},
\bfpage{452}.
\doiurl{10.1086/322405}.
\adsurl{http://esoads.eso.org/abs/2001ApJ...559..452Z}.
\end{barticle}
\endbibitem

\bibitem[\protect\citeauthoryear{{Zhao} and {Dryer}}{2014}]{Zha14}
\begin{barticle}
\bauthor{\bsnm{{Zhao}}, \binits{X.}},
\bauthor{\bsnm{{Dryer}}, \binits{M.}}:
\byear{2014},
\batitle{{Current status of CME/shock arrival time prediction}}.
\bjtitle{Space Weather}
\bvolume{12},
\bfpage{448}.
\doiurl{10.1002/2014SW001060}.
\adsurl{2014SpWea..12..448Z}.
\end{barticle}
\endbibitem

\end{thebibliography}

\end{article} 

\end{document}